\documentclass{aastex6}

\usepackage{color}
\usepackage{natbib}

\newcommand{\ms}{m\ s\ensuremath{^{-1}}}

\newcommand{\vsini}{\ensuremath{V\sin i}}

\newcommand{\rhk}{\ensuremath{\log R'_{hk}}}

\begin{document}

\slugcomment{hires survey paper\_version 33}

\shorttitle{HIRES/Keck Exoplanet Survey}

\shortauthors{Butler et al.}

\title{The LCES HIRES/Keck Precision Radial Velocity Exoplanet Survey}

\author{R. Paul Butler\altaffilmark{1}, Steven S. Vogt\altaffilmark{2}, Gregory Laughlin\altaffilmark{3}, Jennifer A. Burt\altaffilmark{2}, Eugenio J. Rivera\altaffilmark{2}, Mikko Tuomi\altaffilmark{4}, Johanna Teske\altaffilmark{1}, Pamela Arriagada\altaffilmark{1}, Matias Diaz\altaffilmark{5},  Brad Holden\altaffilmark{2}, and Sandy Keiser\altaffilmark{1}}
\altaffiltext{1}{Department of Terrestrial Magnetism, Carnegie Institution for Science, Washington, DC 20015, USA}
\altaffiltext{2}{UCO/Lick Observatory, Department of Astronomy and Astrophysics, University of California at Santa Cruz, Santa Cruz, CA 95064, USA}
\altaffiltext{3}{Department of Astronomy, Yale University, New Haven, CT 06511, USA}
\altaffiltext{4}{University of Hertfordshire, Centre for Astrophysics Research, Science and Technology Research Institute, College Lane, AL10 9AB, Hatfield, UK}
\altaffiltext{5}{Departamento de Astronom\'{i}a, Universidad de Chile, Camino el Observatorio 1515, Casilla 36-D, Las Condes, Santiago, Chile}

\begin{abstract}

We describe a 20-year survey carried out by the Lick-Carnegie Exoplanet Survey Team (LCES), using precision radial velocities from HIRES on the Keck-I telescope to find and characterize extrasolar planetary systems orbiting nearby F, G, K, and M dwarf stars. We provide here 60,949 precision radial velocities for 1,624 stars contained in
that survey. We tabulate a list of 357 significant periodic signals that are of constant period and phase, and not coincident in period and/or phase with stellar activity indices. These signals are thus strongly suggestive of barycentric reflex motion of the star induced by one or more candidate exoplanets in Keplerian motion about the host star. Of these signals, 225 have already been published as planet claims, 60 are classified as significant unpublished planet candidates that await photometric follow-up to rule out activity-related causes, and 54 are also unpublished, but are classified as ``significant'' signals that require confirmation by additional data before rising to classification as planet candidates. Of particular interest is our detection of a candidate planet with $M\sin(i)=3.8\,M_{\oplus}$, and $P=9.9$ days orbiting Lalande 21185, the fourth-closest main sequence star to the Sun. For each of our exoplanetary candidate signals, we provide the period and semi-amplitude of the Keplerian orbital fit, and a likelihood ratio estimate of its statistical significance. We also tabulate 18 Keplerian-like signals that we classify as likely arising from stellar activity.

\end{abstract}

\keywords{Stars: planetary systems}

\section{Introduction}
\label{intro}

In 1994, we initiated an extensive long term search for extrasolar planets around nearby F,G, K, and M dwarf stars using the Keck Observatory HIRES spectrometer on the Keck I telescope atop Mauna Kea. Over the years, the HIRES program has registered some notable successes, including the co-discovery of HD 209458b, the first transiting extrasolar planet \citep{henry2000}, the first discovery of a Neptune-mass planet outside the solar system \citep{but2004}, the first direct mass measurement (without $\sin(i)$ ambiguity) of Gliese 876d, the first super-Earth \citep{rivera05}, and many others. The first decade of the survey's productivity was memorialized in 2008, with a then-complete compendium of the orbital characteristics of nearby exoplanets \citep{cumming2008}. Now, as the survey moves into its third decade, we are electing to publish a catalog of all of the precision Doppler velocity measurements that we have obtained at Keck, with the hope that this data will be of value to the exoplanet community.

To date, this Keck-based precision radial velocity survey of the Lick-Carnegie Exoplanet Survey Team (LCES)  has amassed 60,949 precision radial velocities on a target list of 1,624 stars. Figure 1 shows an H-R diagram of HIPPARCOS stars within 100 pc (purple points), and HIRES/Keck target stars (red points). With the exception of a few transiting or other planet signal cases discovered elsewhere, we have been targeting primarily nearby F, G, K, and M main sequence stars which have been heavily pre-selected for low activity (expected stellar jitter) based on \rhk\ chromospheric emission indices determined from emission reversals in the Fraunhofer H and K lines of Ca II.

\begin{figure}
\includegraphics[angle=0,scale=1.0, trim = 30 0 0 0]{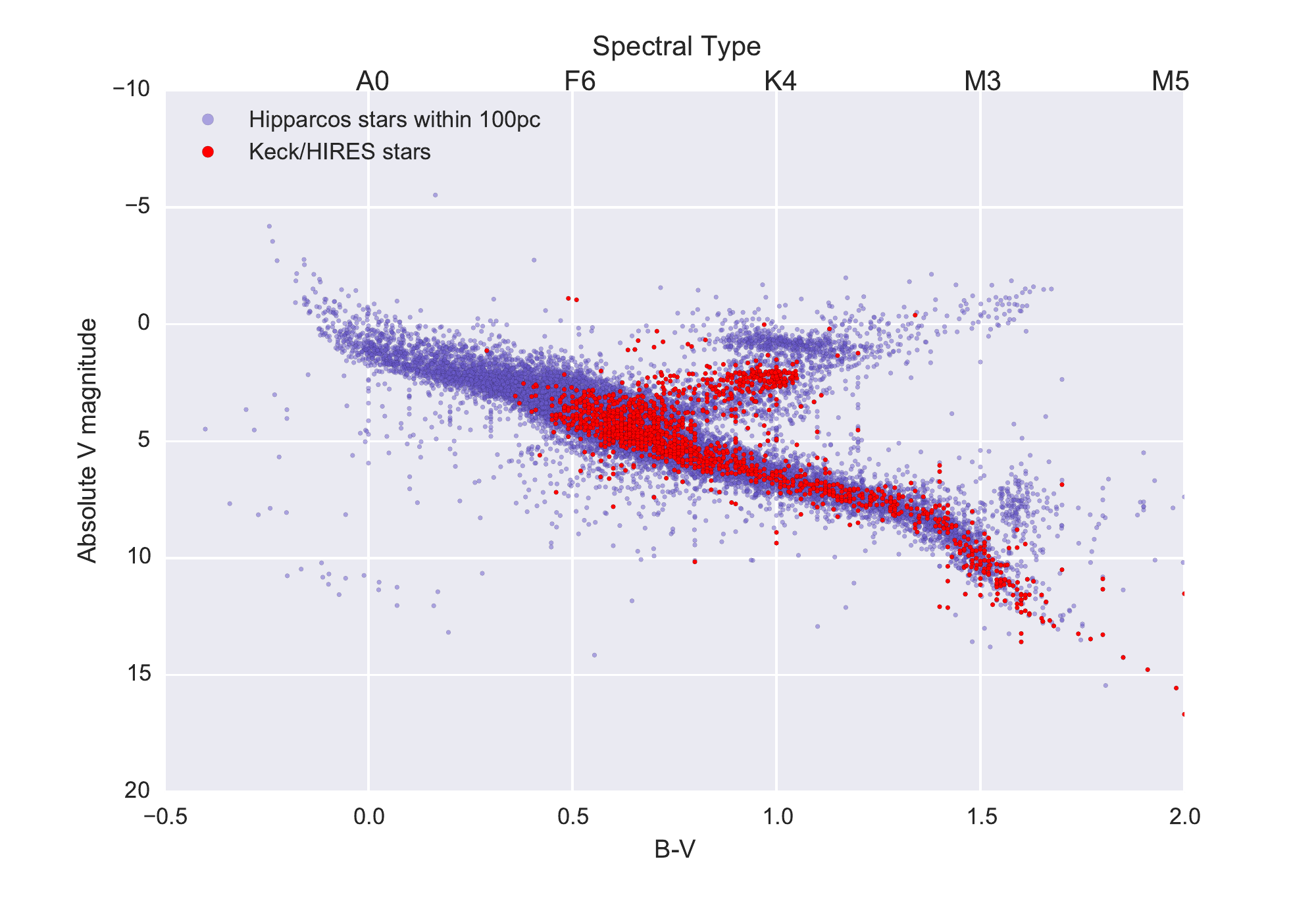}
\caption{H-R diagram of the survey program stars.}
\label{HR diagram}
\end{figure}

\section{Radial velocity observations}

We use the iodine cell technique pioneered by \cite{but96}. All velocities reported herein were obtained with the 
HIRES spectrometer \citep{vog94} at the Keck observatory. 
Radial velocities were measured using an iodine gaseous
absorption cell as a precision velocity reference, placed just ahead of the spectrometer slit in the converging
beam from the telescope as described in detail by \cite{but96}. 
The iodine gas in this absorption cell superimposes 
a rich forest of iodine lines on the stellar spectrum, providing a 
wavelength calibration and wavelength dependent proxy for the point spread function (PSF) 
of the spectrometer. The iodine cell is sealed and temperature-controlled 
to 50.0 $\pm$0.1$^{\circ}$ C, so that the column density of iodine remains 
constant over decades.

For this survey, the HIRES spectrometer was configured to operate at a nominal spectral resolving power 
of $R \sim 60,000$ and wavelength range of 3700\,--\,8000 \AA, however
only the region from 5000\,--\,6200 \AA\, (with iodine lines) was
used in the Doppler analysis. Doppler shifts from the spectra are determined with
the spectral synthesis technique described by \cite{but96}.
For this velocity analysis, the iodine region of the echelle spectrum was subdivided into $\sim$700 
wavelength chunks of 2 \AA\,  each. Each chunk provided an independent measure of 
the wavelength, PSF, and Doppler shift. The final measured velocity 
is the weighted mean of the velocities of the individual chunks. The final uncertainty on each velocity is the standard deviation of all 700 chunk velocities about that mean. Also derived are two chromospheric activity indices. The first is the well-known S-index, obtained from measurement of the emission reversal at the cores of the Fraunhofer H and K lines of Ca II at 3968\AA\ and 3934\AA\ respectively. The second activity index, herein called the H-index, is a measure of the chromospheric emission component at the H$\alpha$ Balmer line of hydrogen at 6563\AA\ \citep{Gomes2011}.

HIRES was neither originally intended, nor specifically optimized for, extreme precision radial velocity work. Rather, it was optimized as a general-purpose instrument,  to provide moderately high resolution (50,000-70,000) echelle-style spectra with good sky background subtraction on faint targets (V$\sim18-22$) such as quasars. In the latter stages of HIRES assembly and test, however, we added an iodine gaseous absorption cell to facilitate precision radial velocity work. We drew on a combination of HIRES commissioning time and early allocations of UC-Keck nights to develop a precision velocity data reduction pipeline, in anticipation that HIRES could be employed to facilitate exoplanet discovery. When NASA secured a 1/6 share of Keck and issued calls for proposals to find exoplanets, we were able to garner enough nights on the heavily-shared Keck facility to mount a significant exoplanet discovery program at Keck, combining all of our UC-Keck nights with those from the NASA-Keck TAC. While this program was remarkably successful from 1995-2004, the original Tektronix/SITE 2K x 2K 2048 CCD was actually an engineering-grade backup device from the LRIS instrument, and it limited the radial velocity precision achievable with HIRES. Chief among its drawbacks were comparatively large (24-micron) pixels, a non-flat (convex) focal plane, and a relatively poor and non-linear Charge Transfer Efficiency (CTE).

\begin{figure}
\includegraphics[angle=0,scale=0.65,trim=40 0 0 0]{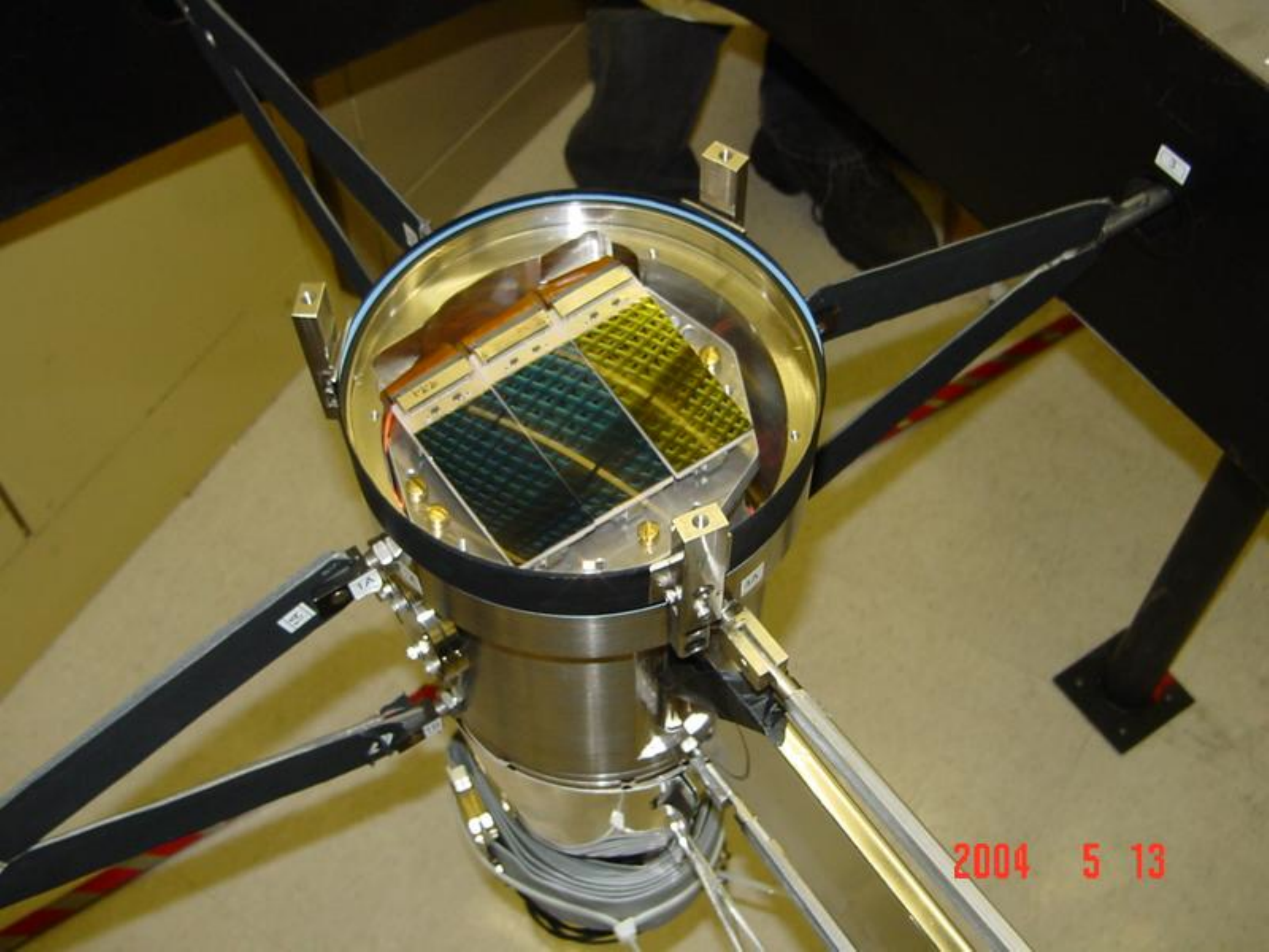}
\caption{HIRES 3-CCD mosaic focal plane upgrade of August 2004}
\label{(B-V) histogram}
\end{figure}

In August-2004, with partial NASA funding, we performed a major upgrade to HIRES. The Tektronix/SITE CCD was replaced with a 3-chip mosaic of MIT-Lincoln Lab 2K x 4K CCD's. These MIT/LL devices feature higher quantum efficiency (QE), smaller (15-micron) pixels, higher and more linear CTE, and a flat imaging surface. At the same time, we remade the final optical element of the HIRES super-camera \citep{epp93} to take full advantage of the now-flattened imager surface. Figure 2 shows a view of the focal plane upgrade (with its internal scattered light baffle removed to better show the CCDs) that we installed in August 2004. The bottom, middle, and top CCD's were optimized respectively for highest QE in the UV, visible, and near-IR. The upgrade was done such that all iodine lines fall onto the middle CCD of the 3-chip mosaic, simplifying data reduction. Also visible in Figure 2 is the new fused silica lens that serves as the dewar window, and flattens the focal plane to match the flat CCD mosaic. The result of the upgrade was a marked improvement in limiting velocity precision, from $\sim$ 4-5 \ms\ to 1-2 \ms. We refer hereafter to velocities obtained prior to August 2004 as ``pre-fix" velocities, and velocities obtained after August 2004 as ``post-fix" velocities. We find no significant velocity offset in our data reduction pipeline between pre-fix and post-fix velocities, and thus do not invoke any velocity offset parameter between pre- and post-fix data in our analyses.

Typical exposure times for target stars varied from a few minutes each, to a maximum of 10 minutes. In general, exposure times were made long enough to provide some averaging over low-degree stellar p-modes, which exhibit typical time scales of $\sim5$ minutes on G dwarfs \citep{Goldreich77}. In cases where exposure times were shorter than 5 minutes, multiple sequential exposures were averaged to smooth over the p-modes. Except for special cases involving high cadence issues, in general, each target was observed only once per night. Maximum exposure times were capped at 10 minutes to avoid introducing additional error sources from increased incidence of cosmic rays, and from increased uncertainty in the determination of the photon-weighted mid-point of the exposure (which determines the barycentric correction). Each minute of error in the exposure midpoint produces a 2 \ms\ error in the barycentric correction. This correction depends on Hour Angle and Declination as well. To optimize exoplanet detection, one must determine the true time centroid of an observation  to a level of $\pm\ 15$ seconds to render barycentric correction error insignificant relative to an overall error budget totaling of order 1 \ms\ .

In 1997, to aid the exposure centroiding, we commissioned an exposure meter for HIRES that uses a propeller mirror to pick off a small fraction ($\sim1\%$) of starlight that has passed through the spectrometer slit and samples that light at a 1\ Hz rate. Upon completion of the exposure, a photon-weighted midpoint for the exposure is calculated and used as the time mark for the barycentric correction \citep{kib06}. The exposure meter also is routinely used to terminate the exposure at any desired S/N level.

\section{Observing Target List Basic Properties}

Figure 3 shows a histogram of the apparent V magnitudes of the stars in our target list. We generally avoided observing stars brighter than V $\sim5$ at Keck as these were easily done with the Hamilton spectrometer \citep{Vogt1987} on the Lick Shane 3-m. And, to ensure at least some coverage of all targets on our large target list, we kept our individual exposures under 10 minutes maximum, which led to a faint magnitude limit of about V=14. The median apparent magnitude of the survey's target list is V = 8.0. When we reported velocity measurements in support of planet detections, we averaged the data over 2-hour bins. For the data presented in this paper, however, we present all velocities as unbinned.

\begin{figure}
\includegraphics[angle=0,scale=1.2,trim=0 0 0 0]{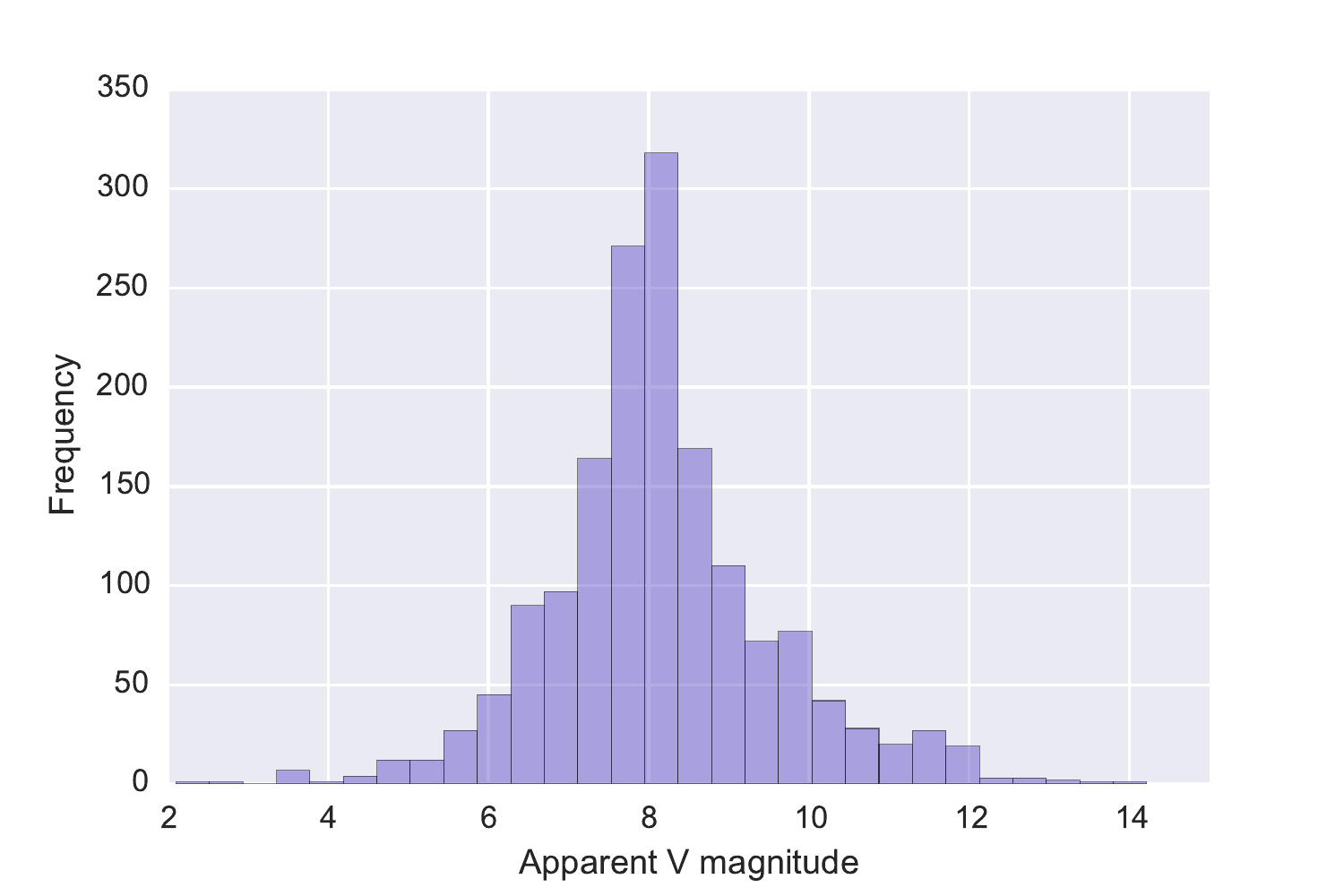}
\caption{Histogram of target star apparent V-band magnitudes}
\label{V magnitude histogram}
\end{figure}

Figure 4 shows a histogram of the (B-V) colors of the stars in our target list. Our target list MK spectral types run from about F5V to M6V. We opted to largely forego surveying earlier than mid-F spectral types, due to decreasing stellar line density and increasing propensity for Delta Scuti type variations \citep{Breger2000}. Such oscillations cause low-level quasi-periodic RV variations that can mimic Keplerian signals. Our latest spectral types are around M5.5V - M6V at the faint end limit.

\begin{figure}
\includegraphics[angle=0,scale=1.2,trim=0 0 0 0]{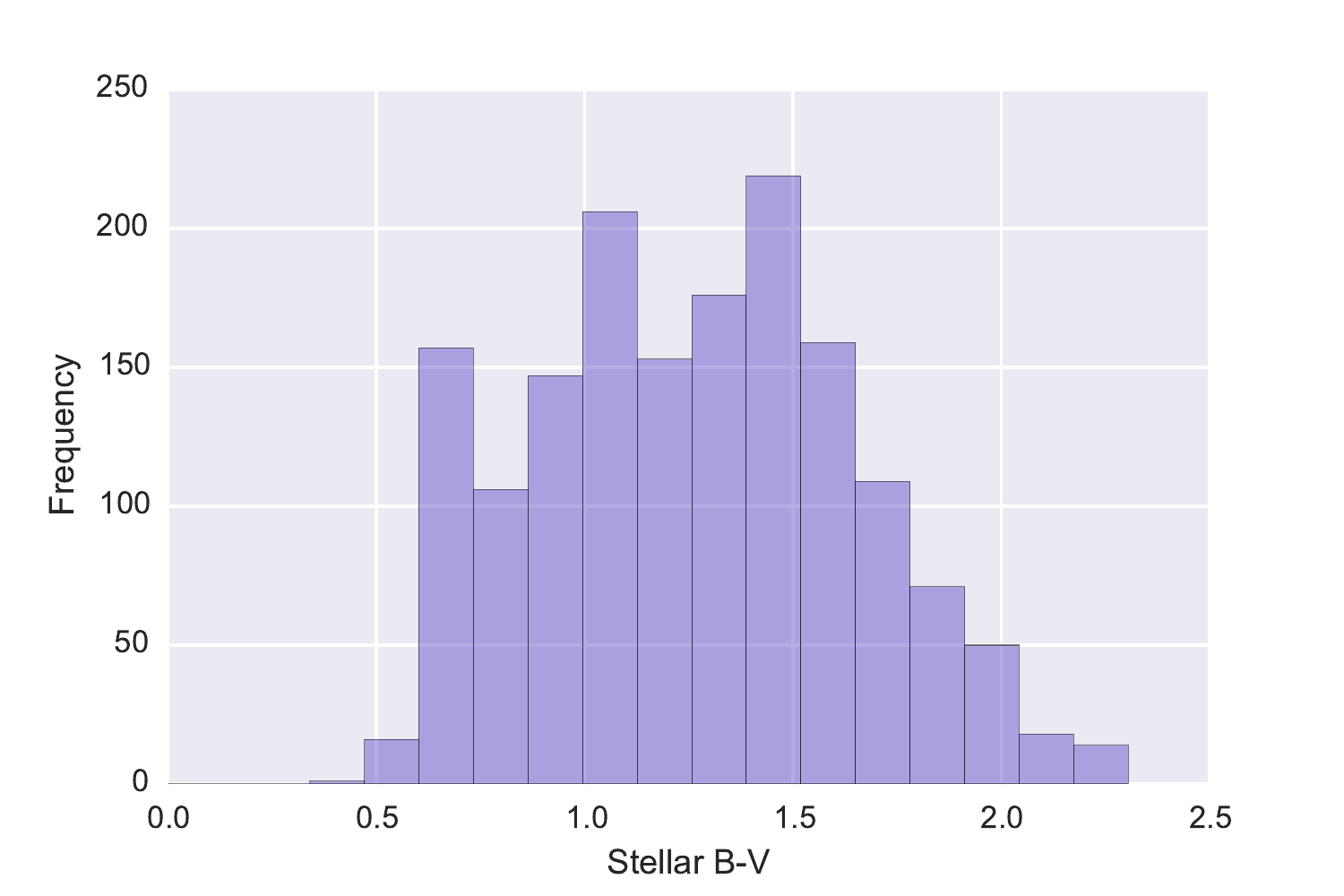}
\caption{Histogram of target star (B-V) colors}
\label{(B-V) histogram}
\end{figure}

Figure 5 shows histograms of the internal uncertainties (pink) of all the radial velocities and the expected stellar jitter (blue) for all program stars. Each reported radial velocity is the average of the individual velocities extracted from $\sim700$ spectral chunks across the iodine region of the echelle spectral format. The internal uncertainty for each velocity is simply the standard deviation of those velocities about that average. As such, it does not include any other unknown systematic errors that may be present in the reduction pipeline. The external error associated with any velocity will generally be larger. To maximize chances of exoplanet detection, we used knowledge of the chromospheric activity S-value to limit the survey primarily to the intrinsically quietest stars at any spectral type. In general, as can be seen from Figure 5, we limited our survey to stars with expected stellar jitter below about 5 \ms.

Our standard metric of chromospheric activity in target stars is the Mt.~Wilson ``S-value". The S-value quantifies the ratio of flux from 1 \AA\ bins located at the centers of the Ca II Fraunhofer H \& K lines, as compared to two broader bandpasses lying 250\AA\ to either side of these lines \citep{Duncan1991}. We computed this ratio for every spectrum taken of each of the stars in the HIRES data sample using the procedure described in \cite{Wright2004}. The results are then calibrated to standard Mt.~Wilson S-values using those stars that overlap between our sample and the original Mt.~Wilson survey. We then used the (B-V) color of each star along with the median of its calibrated S-values to calculate its expected jitter (blue histogram in Figure 5) following the steps outlined in \cite{Isaacson2010}. The median expected stellar jitter of our sample is 2.13 \ms, while the median of our internal uncertainties is somewhat smaller at 1.56 \ms\ and thus well-matched to the expected stellar jitter levels from the sample.

\begin{figure}
\includegraphics[angle=0,scale=1.0,trim=20 20 0 0]{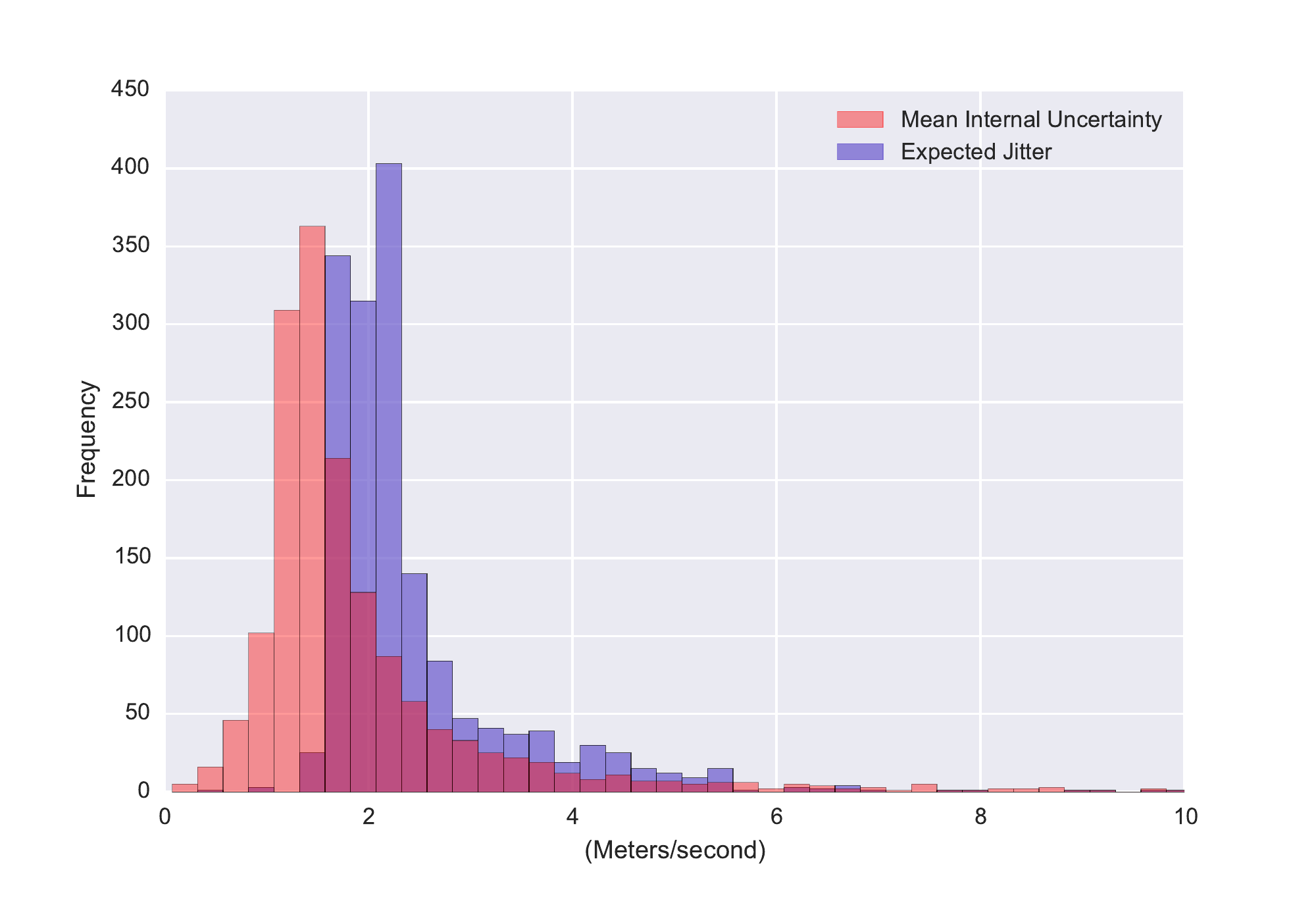}
\caption{Histogram of velocity internal uncertainties and expected stellar jitter}
\label{uncertainties/jitter histogram}
\end{figure}

Figure 6 shows a histogram of distances (from HIPPARCOS parallaxes) of the program stars out to 150 pc. The survey is heavily biased toward the nearest stars, especially the nearest K and M dwarfs. The median distance of our program stars is 36 parsecs. Figure 7 shows a cumulative distribution of observation time baselines for all program stars. 80\% of all target stars were observed over a time baseline of at least 1000 days, while 50\% were observed for at least 3000 days. The longest time baselines for stars is 6300 days or about 17 years.

\begin{figure}
\includegraphics[angle=0,scale=1.2,trim=0 10 0 0]{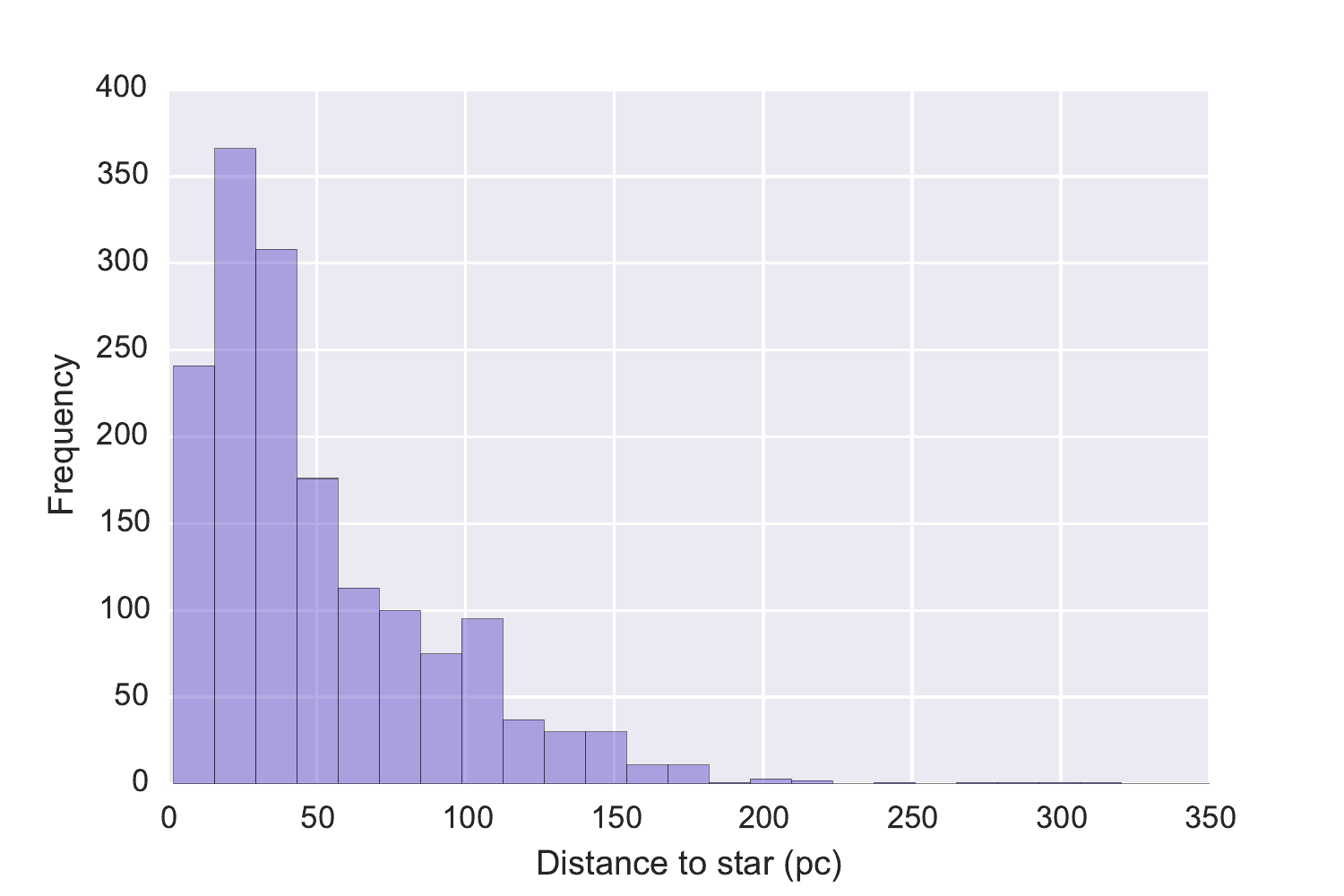}
\caption{Histogram of target star distances}
\label{stellar distance histogram}
\end{figure}

\begin{figure}
\includegraphics[angle=0,scale=0.9,trim=0 10 0 0]{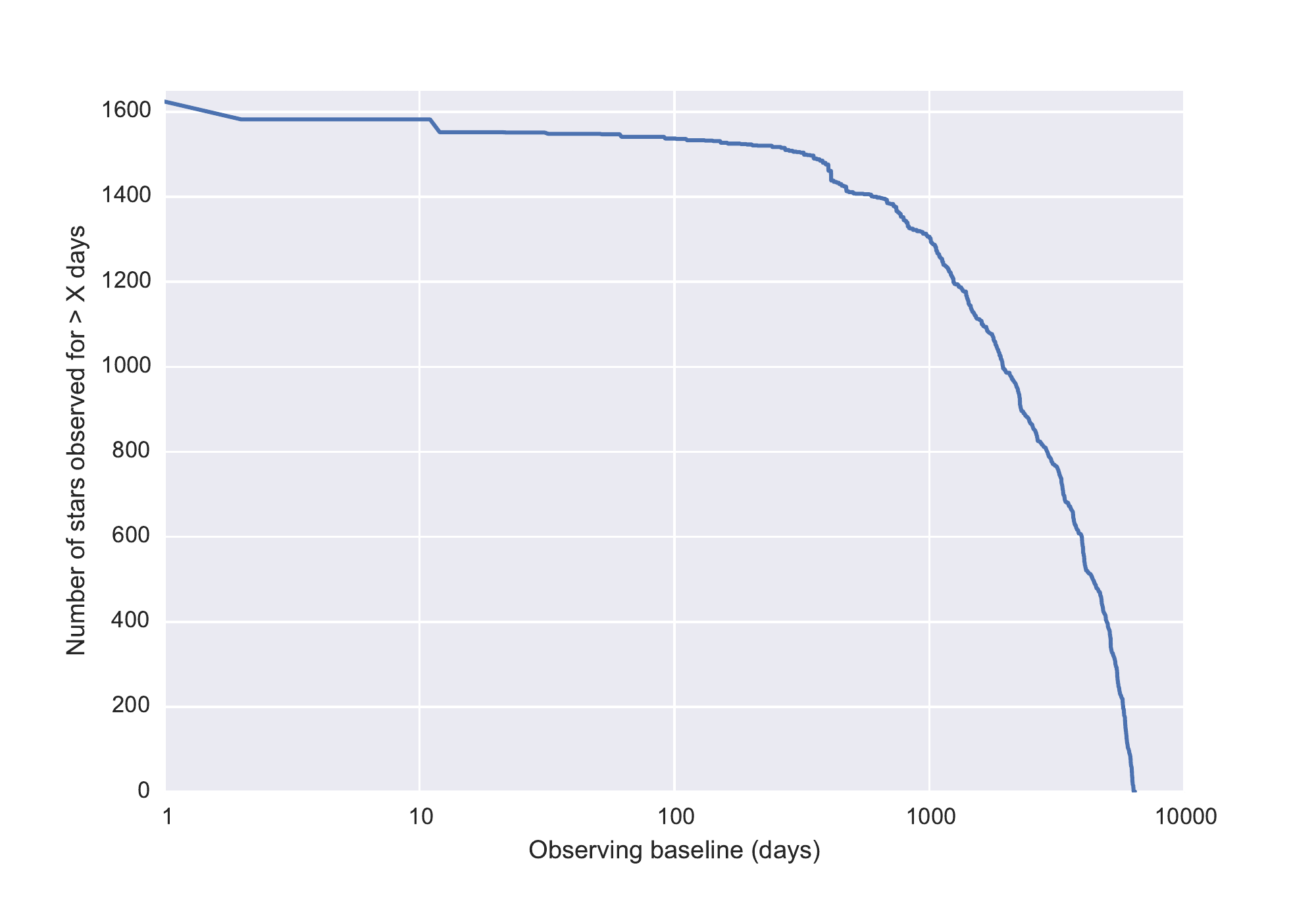}
\caption{Cumulative distribution of observing time baselines}
\label{time baseline distribution}
\end{figure}

Figure 8 shows a histogram of the spectral resolution obtained for all survey spectra. As mentioned above, HIRES was designed primarily for moderately-high spectral resolution on very faint objects, and thus is not ideally suited to extreme precision radial velocities. As shown by \cite{Bouchy01}, the radial velocity information content or ``quality factor", $Q$ of a stellar spectrum depends on spectral resolution. For resolutions lower than $R<50,000$, most of the lines are blended and $Q$ scales roughly linearly with $R$. For resolutions, $R>50,000$, the lines become better resolved and dependence on R flattens out with the slope of the relation linearly decreasing as 1/R. HIRES has a {\it Throughput} (a figure of merit for grating spectrometers defined as the product of spectral resolution times slit width in arc-seconds) of about 39,000 arc-seconds. This value is geometrically fixed by the blaze angle of the echelle, the collimated beam diameter, and the telescope's primary mirror diameter. At that {\it Throughput}, the 0.861 arc-second nominal slit guarantees that the resolution is never less than 45,300 (no matter how over-filled the slit becomes from poor seeing). However, if the seeing is better than 0.861 arc-seconds, the spectral resolution for any exposure will be higher, especially if the guiding is stable and the exposure is short.

Unlike fiber-fed approaches such as that pioneered by HARPS \citep{pepe2000}, where the seeing disk overfills the fiber aperture, and the fiber itself provides a high degree of image scrambling, thus assuring a stable width point spread function input to the spectrometer, HIRES uses a simple long slit, without any scrambling. The advantage of the long slit is less light lost at the slit, since the slit is typically much longer than the seeing disk's full-width at half-maximum (FWHM), and thus light lost at the long slit scales only as slit width, rather than as the square of the diameter of the fiber for the fiber-fed case. However, the lack of image scrambling and its attendant point spread function (PSF) stabilization requires that the PSF be solved for explicitly as part of the RV reductions. The iodine cell's thousands of unresolved absorption lines serve as the proxy for the PSF, allowing variations of the PSF from one observation to the next to be adequately modeled and removed from the analysis so as not to contribute systematic errors in velocity.

At the same time, with HIRES, the FWHM of the seeing at Keck is often less than the width of our standard 0.861 arc-second slit, in which case the spectral resolution becomes a function of the seeing. For cases of very bad seeing, when the seeing disk mostly overfills the slit, resolution could, in principle, degrade to the value 39,000/0.861 set by the {\it Throughput}, or to about 45,000. In practice, we rarely see resolutions less than $R\sim49,000$. When the seeing is extremely good, and the exposure is short, resolutions as high as $R=85,000$ can be achieved, corresponding to an implied seeing FWHM of about 0.46 arc-seconds. The mean resolution across the 60,949 velocities of the survey is $\sim$60,000 as shown in Figure 8, and implies a mean seeing at Keck of about 0.65 arc-seconds over the $\sim$20-year survey.

\begin{figure}
\includegraphics[angle=0,scale=0.9,trim=0 20 0 0]{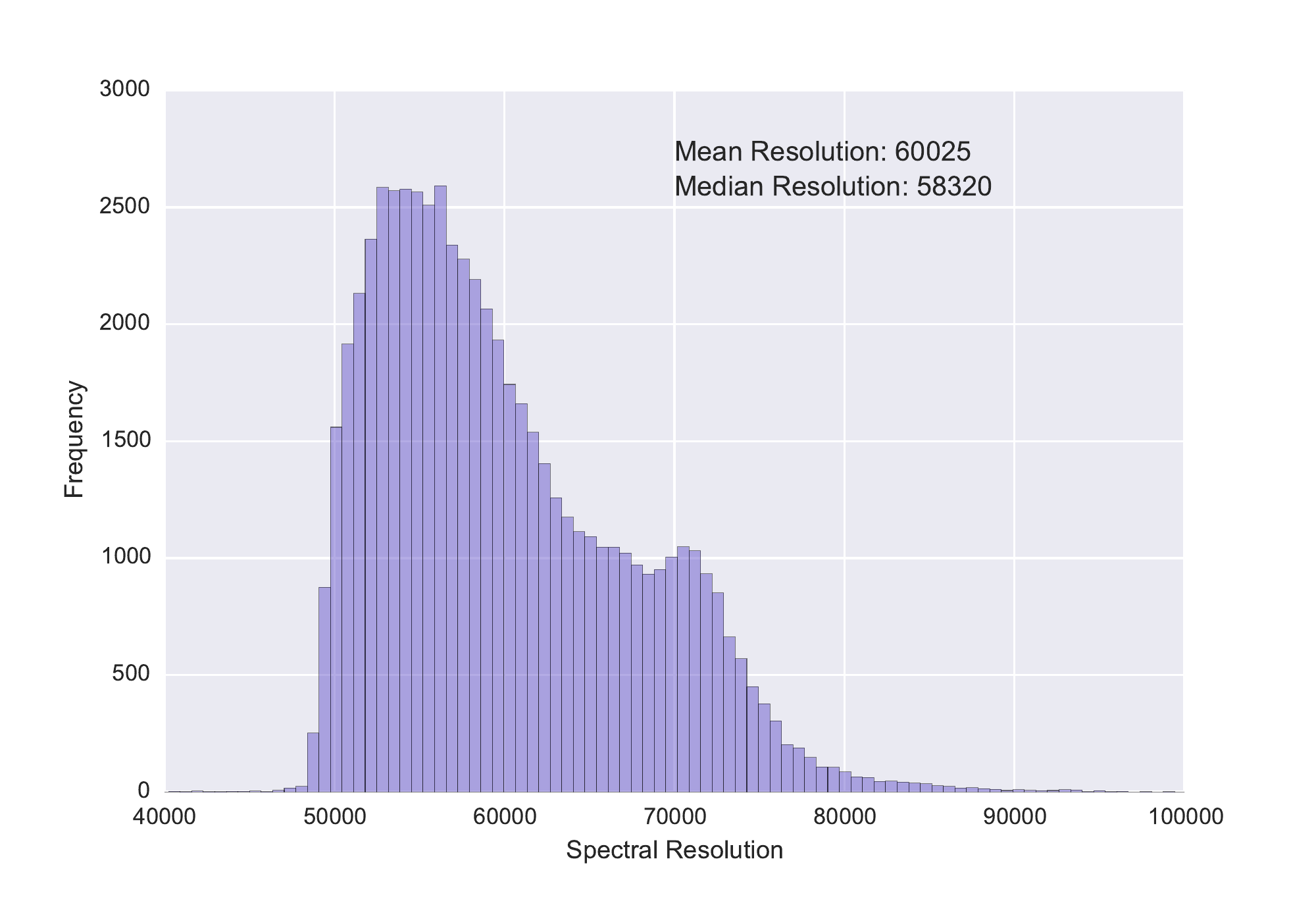}
\caption{Histogram of spectral resolutions of all survey spectra}
\label{resolution histogram}
\end{figure}

Figure 9 shows a time history of the spectral resolutions in this survey. The survey started out using Decker B1 which has a width of 0.57 arc-seconds. On JD 2451983 (March 14, 2001) we mostly transitioned to using either Deckers B5 or C1, both of which have widths of 0.86 arc-seconds. As a consequence, the secondary peak near resolutions of about 72,000 in the histogram of Figure 8 arises from these $\sim$ 7000 observations taken before 2001 with the narrower slit.

\begin{figure}
\includegraphics[angle=0,scale=0.9,trim=0 10 0 0]{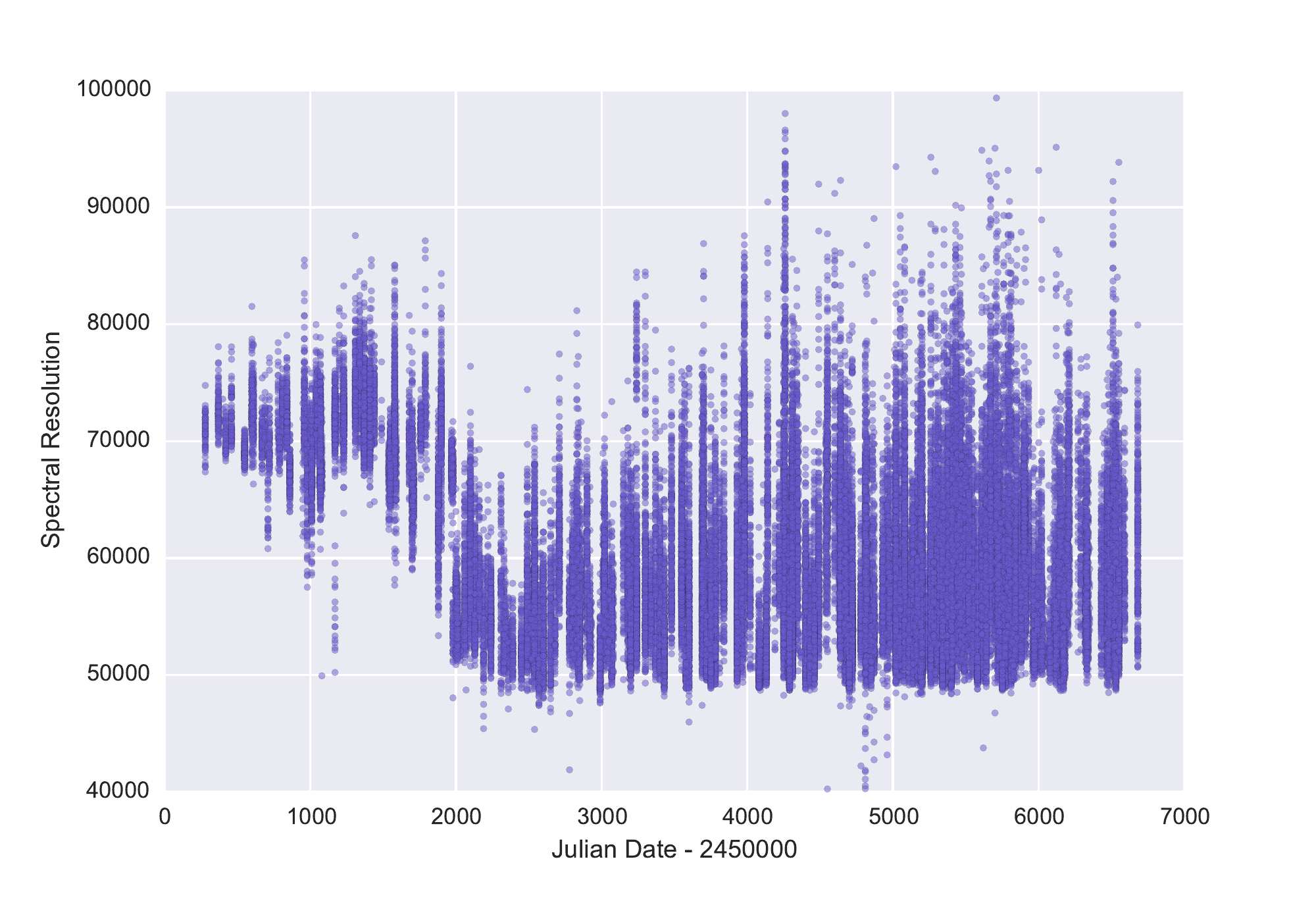}
\caption{Spectral resolutions of all survey spectra vs. Julian date}
\label{resolution history}
\end{figure}

One additional consequence of not using an image scrambling fiber, and having a slit that is typically under-filled by the seeing disk, is that the achieved spectral resolution is also a function of air mass. Increasing air mass generally causes an increase in the seeing disk FWHM, and also an increase in exposure times, resulting in more motion of the seeing disk on the slit during the exposure. Figure 10 shows the dependence of achieved spectral resolution vs. elevation of the star at time of observation. For elevations above $30^{\circ}$, the resolution typically ranges from 49,000 to 75,000 depending on the seeing quality. Below elevations of $30^{\circ}$, the typical resolution degrades steadily with decreasing elevation (increasing air mass).

\begin{figure}
\includegraphics[angle=0,scale=1.2,trim=-10 10 0 0]{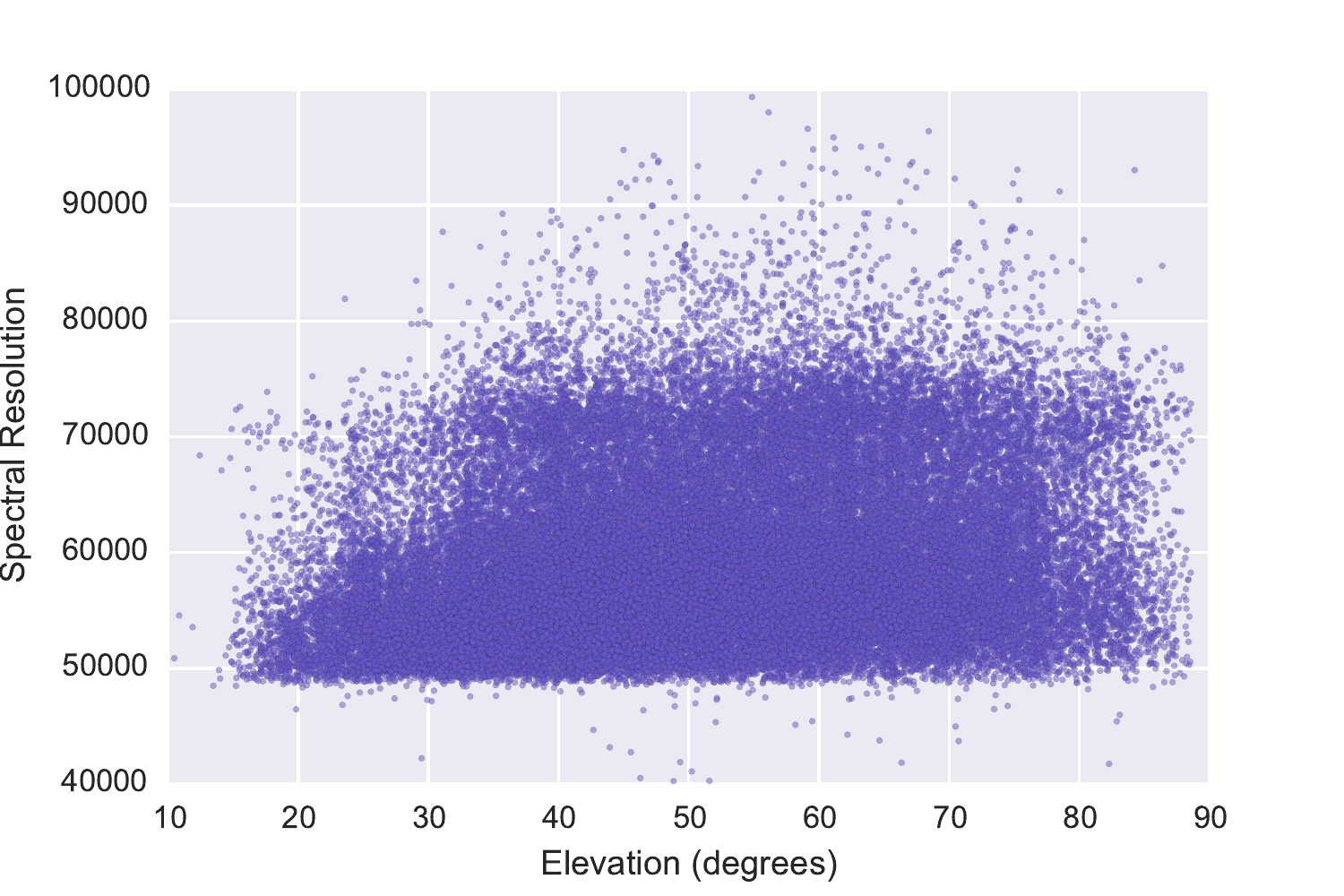}
\vskip 0.1in
\caption{Spectral resolution vs. telescope elevation for all survey spectra}
\label{resolution_elevation}
\end{figure}

Spectral resolution also affects the mean internal uncertainty, which represents the limiting precision of any single observation. This is shown in Figure 11, which plots the mean internal uncertainty of each velocity vs. the resolution of the spectrum from which the velocity was extracted. At  our median resolution of about 58,000, the internal uncertainties are typically 0.8 \ms $<\sigma< 5$ \ms\ (dependent also on stellar spectral type and stellar \vsini). But for resolutions above 75,000, a large fraction of the median internal uncertainties are in the 1 - 2 \ms\ range. Extrapolating that distribution to the R = 115,000 typical of HARPS and its variants implies that per observation median internal uncertainties of about 1 \ms\ could be achieved with HIRES if its resolution could be increased. This can be done straightforwardly by simply narrowing the slit, albeit at the expense of substantial light loss.

\begin{figure}
[b!]
\includegraphics[angle=0,scale=1.2,trim=0 10 0 0]{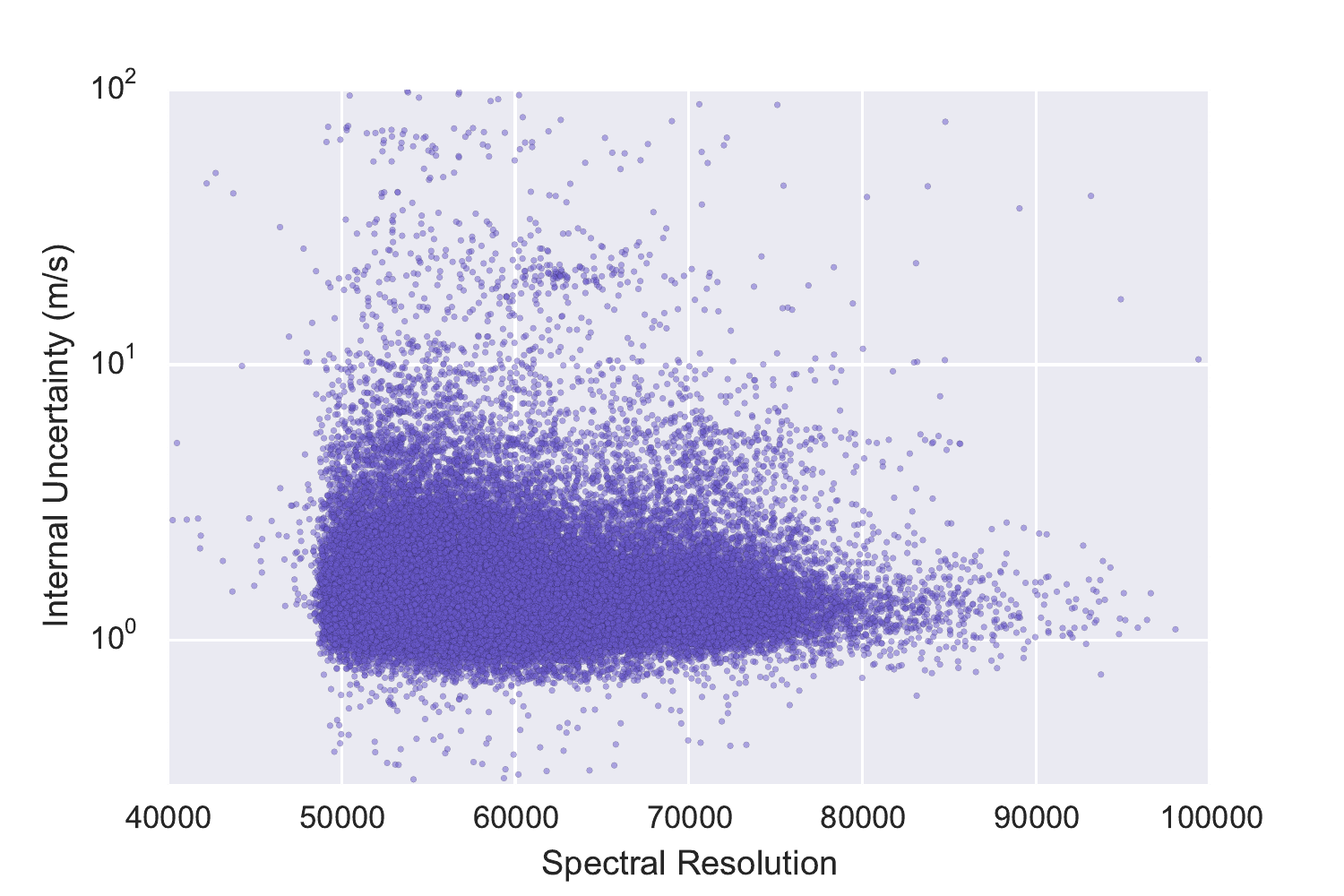}
\vskip 0.2 in
\caption{Mean internal uncertainty vs. spectral resolution for all survey spectra}
\label{intunc vs res}
\end{figure}

When comparing the effectiveness of HIRES for precision radial velocity work against facilities such as HARPS (that were explicitly optimized for such work at a resolution of 115,000), HIRES is at a distinct disadvantage by at least a factor of two because of its factor of approximately two lower spectral resolution. HARPS is optimized for a much smaller 3.6-m telescope, and, even so, suffers the loss of typically 50\% of the light at its 1 arc-second diameter fiber in nominal seeing. Mounting a similar-sized HARPS-like instrument on a telescope as large as the Keck 10-m would mean even more serious light loss at the entrance aperture. Options for improving the {\it Throughput} of HIRES are to invoke adaptive optics (AO), and/or to use an image-slicer. Doing adequate AO in the visible remains technically beyond the immediate near term for HIRES at Keck. But adding an image slicer to HIRES to boost its spectral resolution without losing additional light at the slit should be feasible.

Work at HIRES to improve the spectral resolution for precision radial velocity acquisition has been reported by \citep{spronck15, spronck12}. These authors equipped HIRES first with an image-sliced fiber scrambler feed \citep{spronck12}, and later with a fiber double-scrambler \citep{spronck15}. In the second of the two articles, Spronck et al. report values for their single measurement precision (SMP- equivalent to our internal uncertainty) obtained with the conventional slit-fed HIRES of 1.96 and 2.11 \ms\ for slit-fed observations of two test stars with HIRES. Their SMP on these two stars then improved to 1.49 \ms\ for both stars when HIRES was used with their dual fiber scrambler/slicer. Our survey's median SMP, (Figure 5) obtained from $\sim$61,000 observations, is 1.56 \ms, with 50\% of our velocities attaining significantly higher precisions of 0.5 - 1.0 \ms. From these results we infer that \cite{spronck15} may retain a significant component of noise intrinsic to the data reduction pipeline and/or observing procedure, that is not yet mitigated by the higher spectral resolution and stabilization of the instrument Line Spread Function enabled by the installed fiber scrambler/slicer. Spronck et al's work demonstrates that gains can be obtained from image-slicing and PSF-scrambling HIRES up to a HARPS-like resolution of 120,000 with a fiber scrambler/slicer, but sub-\ms~ measurement precision with this approach has yet to be demonstrated with HIRES. 

\section{Stellar activity indices}

The usual proxy for monitoring stellar activity levels is the S-index and its corresponding derivative \rhk. The S-index is a measure of the emission at the cores of the Fraunhofer H and K lines of singly-ionized calcium due to chromospheric activity. It is often useful in revealing the rotation period of the star, and any longterm activity cycles similar to the Sun's 11-year activity cycle. Chromospheric emission variations due to either or both stellar rotation and longterm activity cycles can produce radial velocity variations that can mimic, and therefore be mistaken for, Keplerian motion of planets. Any periodicity detectable in either broadband photometry or S-index that coincides with a planet candidate's period would cast serious suspicion on the latter's veracity. In some cases, the rotation period of the star can be detected solely from S-index data, even when broadband photometric variations reveal nothing.

We include in this paper, for the first time in such a large collection of HIRES spectra, an additional measure of the stellar
chromospheric activity (hereafter called the H-index) as measured from the H$\alpha$ Balmer line of
hydrogen at 6563\AA. The motivation to use an additional activity
index, besides the S-index, comes from the low flux of M dwarf stars
in the Ca H\&K wavelength region; these stars are intrinsically
brighter at redder wavelengths. Several other authors have also explored how H$\alpha$ flux variations compare to other stellar activity indices, stellar properties, and radial velocity measurements, e.g. \cite{Pasquini1991}, \cite{Gomes2011}, \cite{Robertson2013}, and \cite{Gomes2014}. We can build on their work with a much larger sample that is both uniform (all from HIRES) and spanning a wide spectral type range (late F to early M). Since this index is new to HIRES data, a brief description of its calculation and use is given here. A more
detailed discussion of the HIRES H-index data is in preparation (Teske et al. 2016, in preparation).

Similar to the S-index, the H-index quantifies the amount of flux
within the H$\alpha$ line core compared to the local continuum. We use
the \cite{Gomes2011} prescription, which defines the
H-index as the ratio of the flux within $\pm$0.8~\AA\ of the H$\alpha$
line at 6562.808~\AA\ to the combined flux of two broader flanking wavelength
regions: 6550.87$\pm$5.375 \AA\ and 6580.31$\pm$4.375 \AA. We derive a wavelength
solution specifically for the HIRES H$\alpha$ echelle order from Th-Ar
lamps (rather than using the extrapolated wavelength solution from the iodine region
of the spectrum) and divide out a simple third-order polynomial fit to
the continuum (excluding the H$\alpha$ line wings).
The wavelength solution is refined in each individual spectrum with cross-correlation against the NSO solar atlas\footnote{The development of the NSO Digital Library has been generously supported by the National Science Foundation  through its National Space Weather
Program, and by NASA under the Upper Atmosphere Research Program.}, rebinned to the resolution of the object spectrum. This is necessary, as we do not a priori have the intrinsic stellar velocity measured for
each spectrum. In a small subset of spectra, this solar atlas cross-correlation does not work properly because the H$\alpha$ line is in emission, or because the signal-to-noise of the spectrum is too low (making the H$\alpha$ absorption feature relatively weak). In these
cases, the object spectrum does not match well to the deep H$\alpha$ absorption feature in the solar spectrum, and this leads to an incorrect wavelength solution. After the cross-correlation step, a second continuum normalization is carried out by dividing the object spectrum by the solar atlas, smoothing this fit, and then multiplying the object spectrum by this smoothed fit. Any wavelength mismatches between the object and solar atlas then create waves in the continuum. These issues cause anomalously large H-index values; examination by eye of individual spectra suggests a cut-off of ``good'' versus ``bad'' H-index values between 0.062 and 0.064. Thus we advise caution when using or drawing any conclusions from H-index values above 0.062; this affects only $\sim$1\% of the stars presented here, most of which are late type. In addition, the H-index is presented only for post-fix HIRES spectra (that were collected after the detector
upgrade in August, 2004), and only when the H$\alpha$ region of the raw
spectrum is not saturated ($\ge$ 52,860 counts, determined empirically).

We show in Figure 12 our median H-index  and S-index measurements for each star,
compared to its (B-V) and (V-K) colors. This is slightly different than the \cite{Gomes2011} and \cite{Robertson2013} results in that we see a clear, though not linear, relationship between H-index and color. We note, however, that these H-index values are not corrected for intrinsic stellar luminosity, so variability of the continuum flux with stellar mass, metallicity, or T$_{eff}$ may affect their interpretation \citep{Robertson2013}. What is clear from Figure 12 is that the H-index correlates much more tightly with stellar color (and temperature) than does the S-index. As most of our program stars were specifically chosen to be either main sequence or only slightly evolved sub-giants,
this correlation also manifests as a strong correlation with absolute V-magnitude in Figure 12. 

In Figure 13 we compare the S-index dispersion
with the H-index dispersion. As expected, as the
dispersion in the S-index for a given star increases, so does the
dispersion in its H-index. The stars with the largest variation in
both indices are late-type (almost exclusively M dwarf) stars.

\begin{figure}
\includegraphics[angle=0,scale=0.70,trim=60 100 0 0]{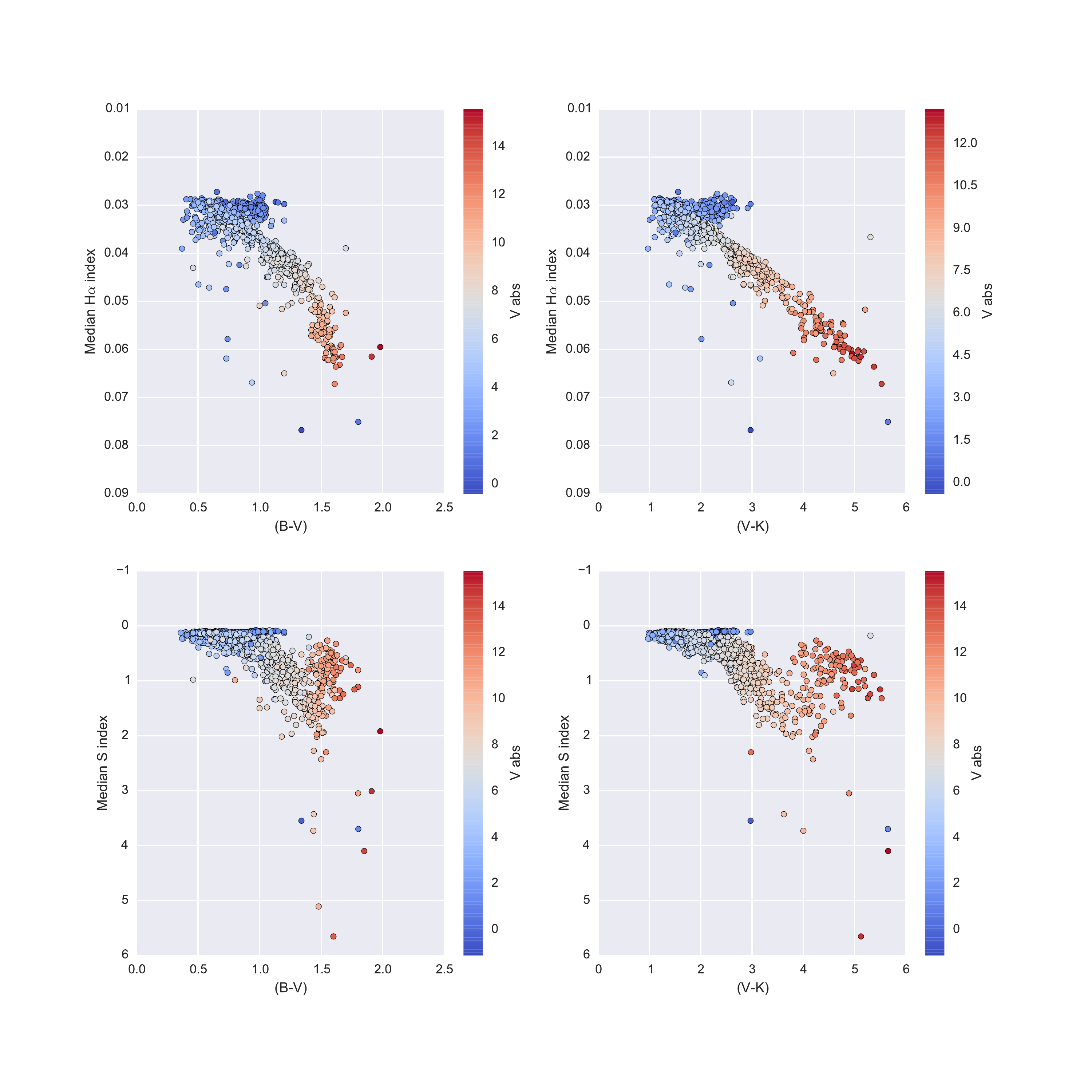}
\vskip 0.25 in
\caption{H-index (top plots) vs. (B-V) color (left), and (V-K) color (right). S-index (bottom plots) vs. (B-V) color (left) and (V-K) color (right). Points are color-coded by stellar absolute V magnitude. Note that H-index values $\geq$0.062 are to be treated with caution, as explained in section 4.}
\label{H-index vs, colors}
\end{figure}

\begin{figure}
[b!]
\includegraphics[angle=0,scale=0.9,trim=10 10 0 0]{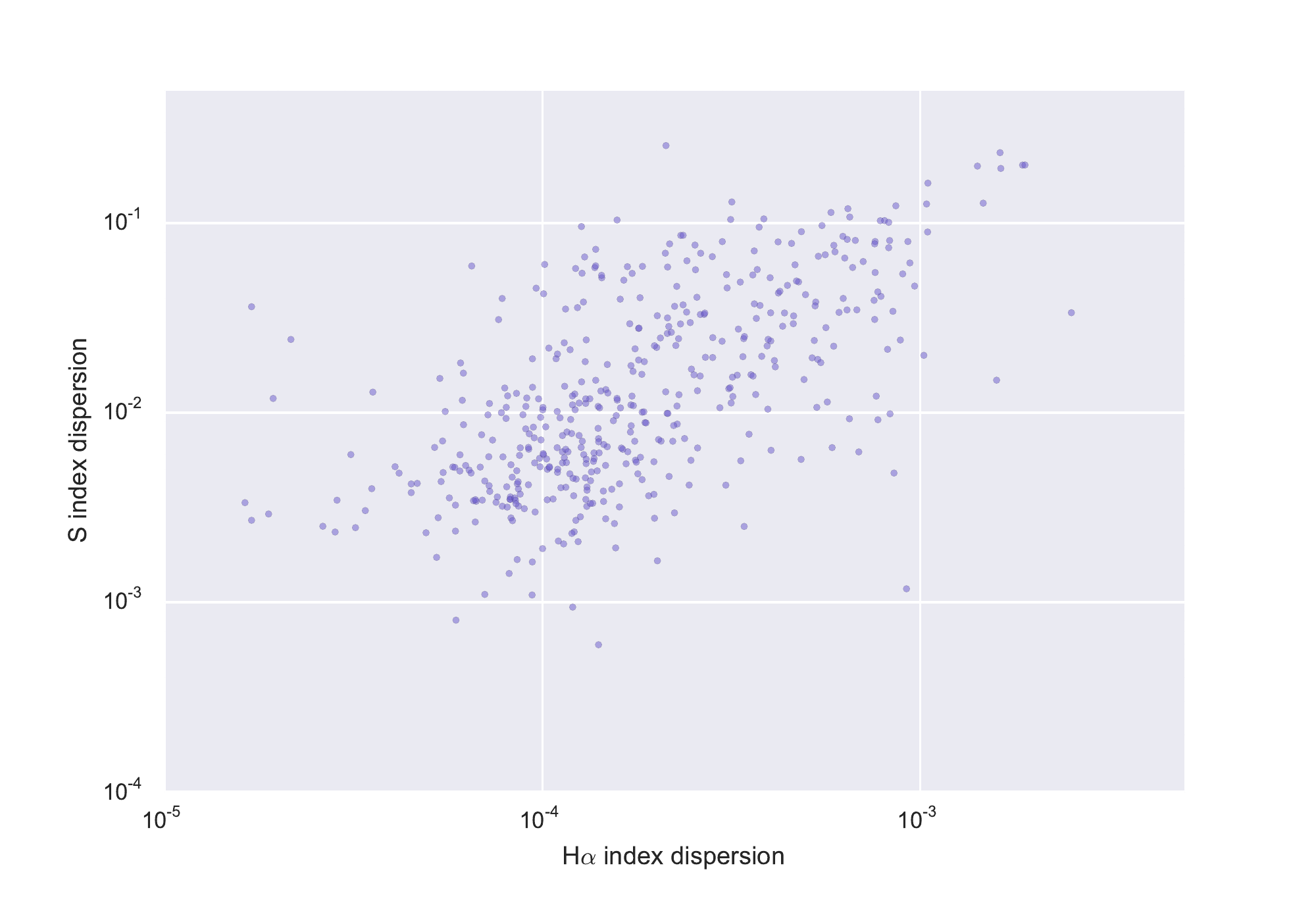}
\caption{HIRES S-index dispersion vs. H-index dispersion}
\label{s-dispersion_vs_h-dispersion}
\end{figure}

\section{Radial velocities and ancillary data}

Table 1 shows a small but representative portion of the full table of radial velocities and ancillary data for all observations of all stars. The full table is available through links in the on-line version of the paper.

\begin{deluxetable}{lcccccc}
\setlength{\tabcolsep}{12pt}
\tabletypesize{\footnotesize}
\tablecaption{Radial velocities and chromospheric activity indices of all program stars. (Note that H-index values $H\geq0.062$ are to be considered with caution, see Section 4 text.) (\textit{Sample: full table in electronic version})
\label{tab:rvdata_LCES/HIRES}}
\tablecolumns{9}
\tablehead{{Target} & {BJD} & {RV [m\,s$^{-1}$]} & {$\sigma$ [m\,s$^{-1}$]} & {S-index} & {H-index} & {Exp Time [sec]} }
\startdata
HD 10002	&	2450462.767	&	-0.53		&	1.09	&	0.151	&	---		&	600  \\
HD 10002	&	2450715.062	&	-6.34		&	1.03	&	0.161	&	---		&	600  \\
HD 10002	&	2450805.761	&	-1.77		&	1.08	&	0.167	&	---		&	400  \\
    ...		&	...			&	...		&	...	&	...		&	...		&	...	\\
HD 10008	&	2453723.831	&	-0.66		&	0.82	&	0.423	&	0.038	&	264  \\
HD 10008	&	2453723.834	&	-2.75		&	0.87	&	0.422	&	0.038	&	235  \\
HD 10008	&	2454129.746	&	9.28		&	1.25	&	0.428	&	0.037	&	81  \\
...		&	...			&	...		&	...	&	...		&	...		&	...   \\
HD 10013	&	2453982.024	&	1.83		&	0.69	&	0.145	&	0.032	&	92  \\
HD 10013	&	2453982.976	&	2.20		&	0.78	&	0.147	&	0.032	&	96  \\
HD 10013	&	2453983.924	&	-2.21		&	0.73	&	0.148	&	0.032	&	92  \\
...		&	...			&	...		&	...	&	...		&	...		&	...   \\
HD 10015	&	2453981.918	&	2846.74	&	0.66	&	0.146	&	0.034	&	101  \\
HD 10015	&	2453982.979	&	2713.58	&	0.66	&	0.142	&	0.034	&	106  \\
HD 10015	&	2453983.917	&	2595.18	&	0.58	&	0.149	&	0.034	&	202  \\
\enddata
\end{deluxetable}

Data presented in Table 1 are (from left to right): Target star name, Barycentric Julian Date, radial velocity (m/s), radial velocity internal uncertainty (m/s), S-index, H-index, and exposure time in seconds. We continue to reduce newly available Keck/HIRES precision velocity data from the NASA/Keck archive.  Newly reduced data is made available on the Earthbound Planet Search web site\footnote{\url{http://home.dtm.ciw.edu/ebps/}}

\section{Planetary candidate signals}

This large $\sim$20-year survey of $\sim$61,000 precision radial velocities contains numerous suspected periodic signals, many of which have already been published over the past two decades by ourselves and/or others as planet detection claims. Some of these remaining unpublished signals are, undoubtedly, evidence of reflex barycentric motion of the star that is attributable to planetary companions in Keplerian motion around their host star. Other confounding periodic signals can arise though from surface inhomogeneities carried into and out of view by stellar rotation, by stellar activity cycles, and by aliasing from other planet signals and observing cadences. Disentangling reflex barycentric motion of the host star from stellar rotation and/or stellar activity effects is a complex task that involves, at the very least, accurate broadband photometry over long time scales. Such photometry is generally not available or even sought for a given star until a promising constant period and phase-stable signal becomes apparent in the RV's.

There is also no uniformly accepted standard across the exoplanet community for false alarm probability (hereafter FAP) determination or other metric that warrants publishability of a candidate planetary signal. Establishing confidence in a given signal requires not only careful scrutiny of stellar rotation, activity effects, and aliasing that can produce spurious detections, but also requires a detailed statistical analysis of the significance of any candidate signal in the data. Such an exhaustive study for all such signals in the data set is beyond the scope of the present paper, and for real utility, such a study would also require gathering all published high-precision Doppler velocities at other telescopes. Rather, we present in Table 2 a list of the most prominent planet-candidate signals that appear in our data sets. Here we show only a representative portion of the full table of signals. The complete version of Table 2 is available through links in the on-line version of the paper. While many of theses signals are undoubtedly caused by planets in Keplerian motion about the host star, issues of potential confusion with stellar rotation and/or jitter will require follow-up work with photometry. In multi-planet cases, detailed modeling of the dynamical stability of the system also need to be verified. No such attempt to do so has been made here.

We performed the analyses of HIRES radial velocity data sets by using an automated Markov chain Monte Carlo (MCMC) posterior sampling algorithm. This algorithm was constructed to both search for signals in the radial velocity time-series and to obtain estimates for Keplerian and nuisance parameters of the statistical model (see Appendix A). We applied the Adaptive Metropolis (AM) MCMC algorithm \citep{haario2001} for parameter estimation and its generalized version, the delayed-rejection adaptive Metropolis (DRAM) algorithm \citep{haario2006} for searches of signals. The analyses proceeded as follows.

First, we performed posterior samplings of a baseline model with $k=0$ (where {\it k} is the number of signals in the model) by applying the AM algorithm to obtain parameter estimates. As in all parameter estimations with MCMC algorithms in the current work, after rejecting $n$ initial chain members (considered to be the burn-in phase in which the chain identified the most probable regions in the parameter space) we divided the remaining chain into three parts and tested non-convergence by applying the Gelman-Rubin statistics for all parameters \citep{gelman2003,ford2006}. If the Gelman-Rubin statistic $R$ in Eq. (25) of \citet{ford2006} was above 1.1 we rejected the chain because it showed evidence in favor of non-convergence and increased the chain length to obtain a more statistically representative sample. Required chain lengths ranged from $10^{5}$ for $k=0$ to $10^{7}$ or more for $k>2$.

We then moved on recursively. Assuming a detection of $k$ signals in the data and that the maximum \emph{a posteriori} (MAP) estimates had been obtained with the AM algorithm, we adopted a model with $k+1$ signals by setting the initial state of the chain such that all but the five Keplerian parameters of the $k+1$th signal were set equal to their MAP estimates as obtained by using the model with only $k$ Keplerian signals. The period parameter of the $k+1$th signal was set to a random value in the set $[P_{\rm min}, T_{\rm obs}]$, where we set $P_{\rm min} = $1 day and $T_{\rm obs}$ is the baseline of the data set. The eccentricity ($e$), and amplitude ($K$) were set equal to zero. With this setting, our model accounted for the k detected signals and enabled the Markov chains to sample the parameter space of the $k+1$th signal in a search for global maxima that could be interpreted as additional signals in the data. This is demonstrated in greater detail below by using the analyses of two data sets as examples (Sections \ref{sec:HD4208} and \ref{sec:HD150706}).

The searches for signals were, however, difficult when there was a very significant Keplerian signal in the data corresponding to e.g. a massive giant planet. In such cases, the Markov chains identified the signal easily but all newly proposed values were rejected because they corresponded to considerably lower posterior values than the MAP estimate in the period space. This stopped the chain from visiting the whole period space, thereby preventing us from seeing whether there were even more significant maxima in other parts of the period space. In such cases, instead of using a likelihood function $L$, we instead used $L^{\beta}$ such that parameter $\beta$ was gradually decreased. If the chains failed to visit the whole period space, we decreased parameter $\beta$ by a factor of 1.1 as long as was necessary to enable the chain to visit all areas\footnote{All areas refers to the chain visiting all 1000 equally long intervals of the $\log P$ space that cover the period space $[\ln P_{\rm min}, \ln T_{\rm obs}]$.} of the period space. We note that, although setting parameter $\beta$ to values below unity changes the likelihood function, the positions of the maxima remain unchanged enabling us to use such tempered samplings for the purpose of searching signals in the noisy radial velocity time-series.

If the sampling of the parameter space of the model with $k+1$ Keplerian signals identified a probability maximum such that the ratio of the two maximum likelihood values $\Delta \ln L_{k+1} = \ln L_{k+1} - \ln L_{k}$ indicated a significant increase with a 0.1\% FAP, corresponding to $\Delta \ln L_{k+1} \geq \alpha = 16.27$, we considered the signal to be significantly detected. In such a case we moved on to estimate the parameters of the model with $k+1$ signals. However, if such a maximum could not be found, we concluded that there is only evidence for $k$ signals in the data (possibly $k=0$).

Table 2 lists all signals in the Keck data sets with likelihood ratio $\Delta \ln L_{j} = \ln L(j=k) - \ln L(j=k-1)$ between models with $k$ and $k-1$ signals for which the ratio exceeds a value of 16.27 corresponding to a 0.1\% FAP threshold of a three-parameter model (amplitude $K$, period $P$, and phase $\phi$). If the likelihood ratio exceeds 20.52 (corresponding to a 0.1\% FAP threshold for a full 5-parameter Keplerian model) the signal is labelled as ``Candidate'' -- otherwise it is called ``signal requiring confirmation'' (SRC). Some signals are interpreted as ``Activity'' according to the criteria presented in the next subsection.

\subsection{Analysis of activity indicators}

We also analyzed the HIRES S-indices in order to assess whether the signals in the radial velocities could be interpreted instead as stellar activity. The S-indices were analyzed simply by calculating likelihood-ratio periodograms for a simple model containing a sinusoidal signal and a linear trend (see Appendix A). We ignored signals with periods below 2 days to avoid detecting daily and yearly aliases whose significance can occasionally exceed the significance of the true signal at longer periods. We applied a 5\% FAP threshold for the detection of signals in S-indices corresponding to a likelihood ratio threshold of $\alpha = 7.82$ for a sinusoidal signal with three parameters.

Table 3 presents a list of estimates of maximum likelihood parameter estimates and standard errors for the ``nuisance'' parameters. $\dot{\gamma}$, $\sigma_{\rm J}$, and $c_{\rm S}$ represent  respectively the linear acceleration, excess white noise or ``jitter'', and the coefficient quantifying the linear dependence of the radial velocities on the S-index. $\Delta \ln L_{\rm S} = \ln L_{\rm S}(k=1) - \ln L_{\rm S}(k=0)$ represents the likelihood ratio statistic of model with a signal ($k=1$) in the S-index with respect to a model without a signal ($k=0$). Only signals exceeding a ratio of 7.82 corresponding to 5\% FAP are shown. $P_{\rm S}$ denotes the period of the signal in the S-index time-series.

\subsection{Interpretation of signals}

The likelihood ratio of $\alpha = 16.27$ is based on the $\chi^{2}$ statistics with three degrees of freedom. We used this limit to search signals but applied a more robust value of 20.52 when considering that they were signals of Keplerian origin. This value is a similar threshold to a 0.1\% FAP (which we have chosen in keeping with the precedent established by \citet{Cumming2004} but for five degrees of freedom), as is the case when comparing models with $k$ and $k+1$ Keplerian signals. We thus interpreted signals exceeding the former threshold as \emph{signals requiring confirmation} (SRC) but those exceeding the latter threshold as \emph{candidate planets} if the following conditions were satisfied.

We consider a periodic radial velocity signal to be a candidate planet if:
\begin{enumerate}
  \item The likelihood ratio $\Delta \ln L_{k}$ exceeds a threshold of $\alpha = 20.52$.
  \item The signal has a well-constrained period and its amplitude is statistically significantly greater than zero.
  \item The signal has no counterparts in the activity indicators.
  \item The data set is not dominated by activity.
\end{enumerate}
Although the first criterion is very clear in its meaning, the others are less so without additional details. 

The second criterion refers to the posterior samplings of the parameter space. Any signal that has an amplitude that is not statistically significantly different from zero (i.e. whose 99\% credibility interval is limited from below by zero) cannot be considered a significant signal even if the first criterion above is satisfied. Similarly, this is also the case when the exact period of the signal cannot be determined reliably due to e.g. poor data sampling and a resulting ambiguity in the signal period. These criteria have been introduced in \citet{tuomi2012} and applied in e.g. \citet{tuomi2014}.

If a given signal with a period $P$ has a counterpart in the activity indicators (S-index or H-index) such that the period of the activity signal $P_{a}$ satisfies $0.8P < P_{a} <1.2P$, we interpret the radial velocity signal as being likely connected to stellar activity (rotation, magnetic cycle, etc.). However, we interpret the signals as candidate planets if they have been reported and interpreted as planetary signals in the literature. Finally, we classify all signals in data sets that are dominated by activity-induced variations as stellar activity. This means that if at least 50\% of the variance in the radial velocities ($\sigma^{2}(m)$) is connected to variations in the S-index, such that $c_{\rm S}\sigma^{2}(S) > 0.5 \sigma^{2}(m)$, we consider the corresponding data set to be dominated by activity-induced variations making the planetary nature of all the corresponding signals doubtful.

\subsection{Example 1. HD 4208 -- a strong signal}\label{sec:HD4208}

HD 4208 is a host to a well-known candidate planet in the Keck sample \citep{butler2006} with an orbital period of 828.0$\pm$8.1 days. This candidate, with a minimum mass of 0.804$\pm$0.073 M$_{\rm Jup}$, gives rise to a radial velocity signal with an amplitude of 19.06$\pm$0.73 ms$^{-1}$ \citep{butler2006}.

\begin{figure}
\center
\includegraphics[angle=0, width=1.1\textwidth,clip]{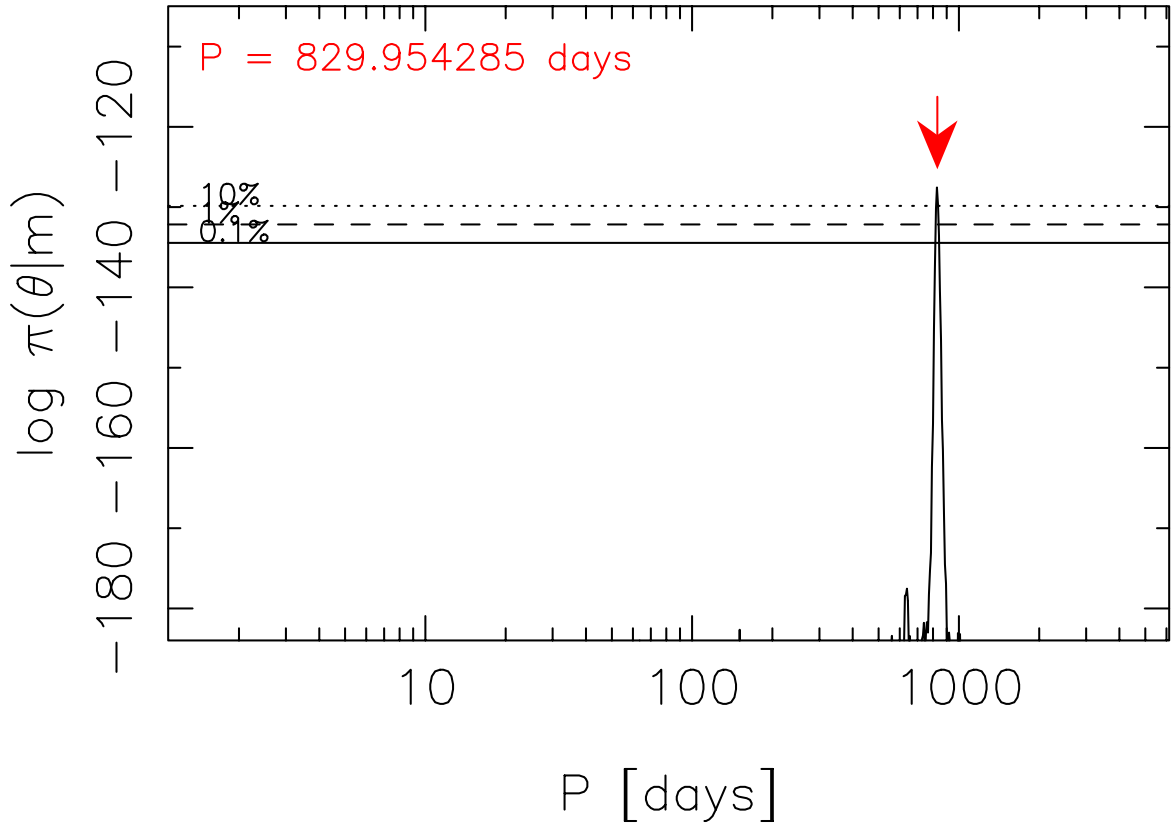}
\caption{Estimated posterior density of the period parameter of the signal in a one-Keplerian model given radial velocities of HD 4208. Red arrow denotes the global probability maximum and the horizontal lines correspond to equiprobability contours at 10\% (dotted), 1\% (dashed), and 0.1\% (solid) of the maximum.}\label{fig:HD4208_posterior}
\end{figure}

Our automatic search for signals identified the periodic signal corresponding to this candidate without complications. During the search, the tempering parameter $\beta$ was decreased to a value of 0.29 before the chains could visit all areas in the period space without getting ``stuck'' to the position of the probability maximum in the period space corresponding to the radial velocity signal of the candidate. As a result, we could obtain an estimated posterior probability density as a function of the period of the signal indicating that this signal is uniquely well-constrained in the period space (Fig. \ref{fig:HD4208_posterior})

\begin{figure}
\center
\includegraphics[angle= 0, width=1.1\textwidth,clip]{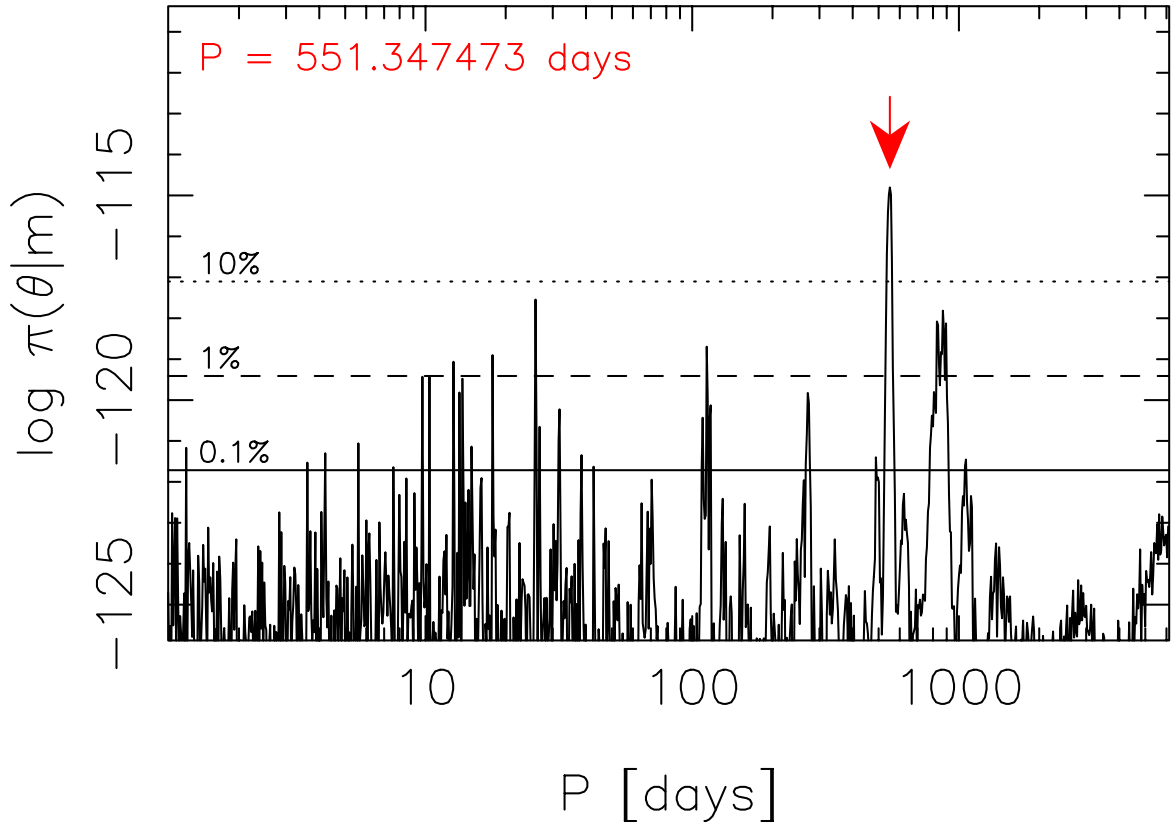}
\caption{As in Fig. \ref{fig:HD4208_posterior} but for the period parameter of the second signal in a model with $k=2$.}\label{fig:HD4208_posterior2}
\end{figure}

As the signal corresponding to the candidate planet reported by \citep{butler2006} was detected clearly and its inclusion in the statistical model yielded an increase in the log-likelihood ratio of $\Delta \ln L_{1} = $62.90, far above the threshold of $\alpha = 20.52$ required for a classification of the signal as a candidate planet, we also increased the number of Keplerian signals in the model to $k=2$ and attempted to find additional signals in the data. This time, the chains did not identify any maxima corresponding to significant signals in the period space and we thus conclude that there is evidence for only one signal in the HD 4208 Keck radial velocities corresponding to the planet candidate first reported in \citet{butler2006}. Our DRAM samplings of the posterior density identified a maximum in the period space at a period of 551 days (Fig. \ref{fig:HD4208_posterior2}) but this maximum was not significant enough to enable classifying it as an SRC and it was accompanied by several local maxima exceeding the 1\% equiprobability threshold, which is a typical outcome in the absence of evidence for additional signals. Although it can be stated that the most prominent period for a second signal in the HD 4208 data is 551 days, the corresponding significance with $\Delta \ln L_{2} =$10.03 is not sufficiently high to conclude that there is evidence in favor of a second signal.

\subsection{Example 2. HD 150706 -- a weak but significant signal}\label{sec:HD150706}

HD 150706 has been reported to be a host to a long-period candidate planet with an orbital period of 5894$^{+5584}_{-1498}$ days and a minimum mass of 2.71$^{+1.14}_{-0.66}$ M$_{\rm Jup}$ giving rise to a radial velocity signal with an amplitude of 31.1$^{+6.3}_{-4.8}$ ms$^{-1}$ \citep{boisse2012}. This detection was based on ELODIE and SOPHIE radial velocities but some of the Keck data was also used to support the detection in \citet{boisse2012}.

With a baseline of only 3970 days, the 58 Keck velocities did not show any evidence in favor of the candidate planet reported by \citep{boisse2012}. Furthermore, we found no evidence for linear acceleration indicative of a long-period companion to the star. The estimated linear acceleration was found to be $\dot{\gamma} = $0.68$\pm$0.55 ms$^{-1}\,$year$^{-1}$, which is not statistically significantly different from zero.

\begin{figure}
\center
\includegraphics[angle= 0, width=1.1\textwidth,clip]{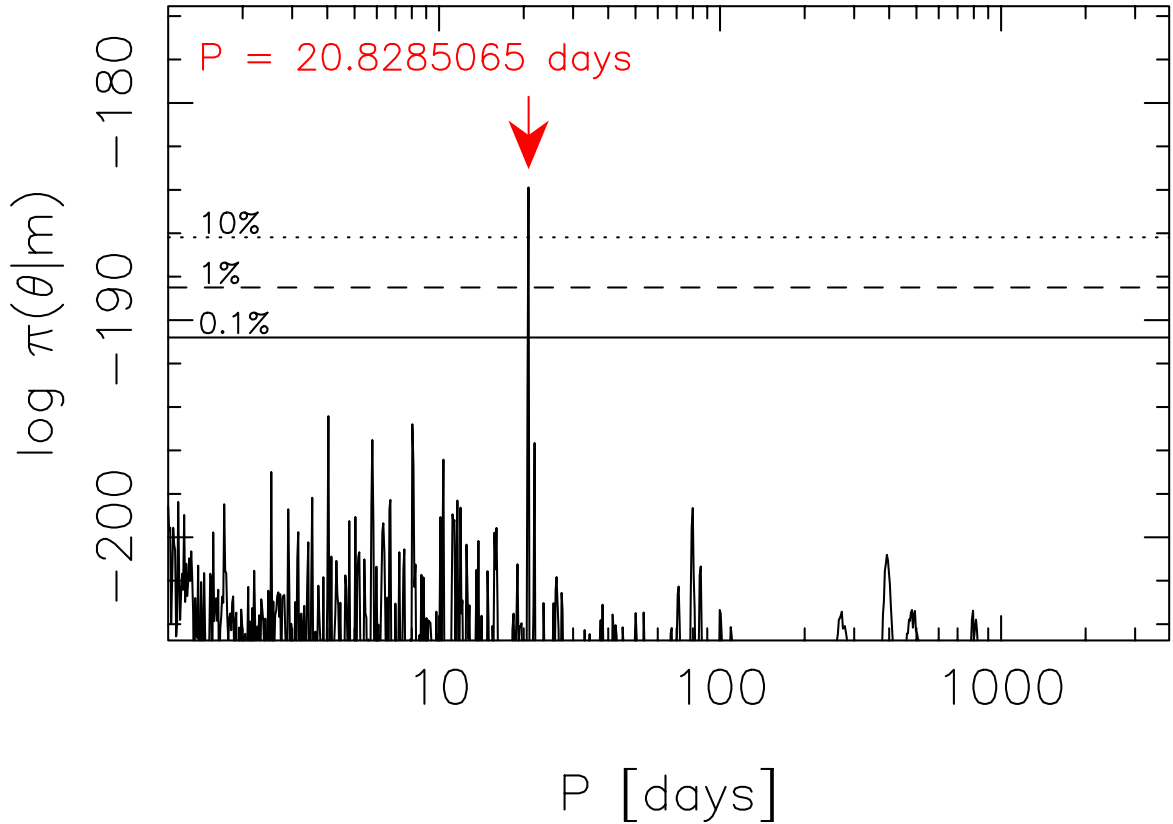}
\caption{As in Fig. \ref{fig:HD4208_posterior} (based on the radial velocities measured for HD 4208) but for HD 150706.}\label{fig:HD150706_posterior}
\end{figure}

\begin{figure}
\center
\includegraphics[angle= 0, width=1.1\textwidth,clip]{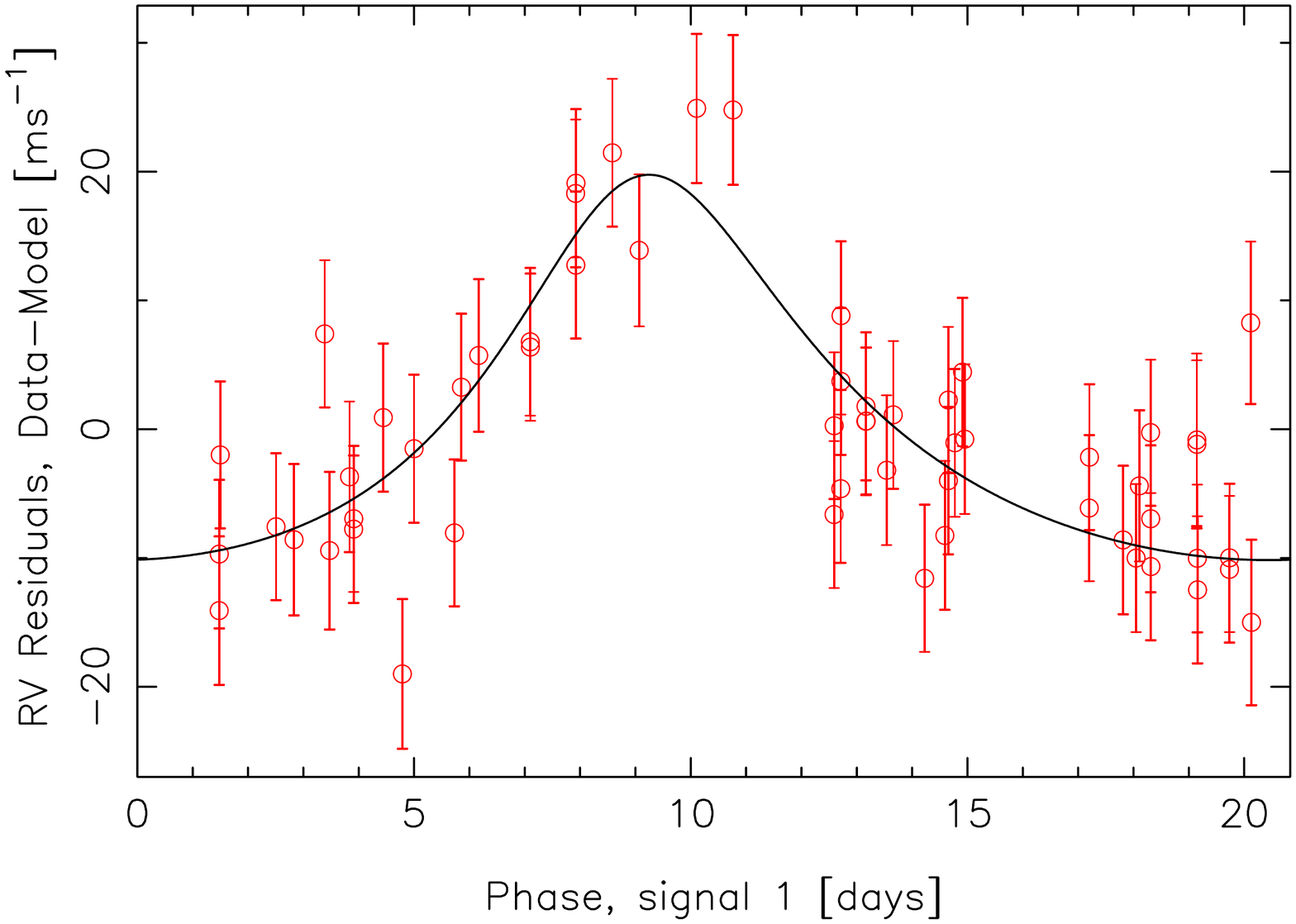}
\caption{Keck radial velocities of HD 150706 folded on the phase of the signal in the data.}\label{fig:HD150706_signal}
\end{figure}

We do, however, find a signal that we classify as a candidate planet orbiting the star at a period of 20.8266$\pm$0.0097 days with an amplitude of 12.55$\pm$2.39 ms$^{-1}$. This signal was uniquely present in the data (Fig. \ref{fig:HD150706_posterior}) and corresponded to a log-likelihood ratio of $\Delta \ln L_{1} =$ 22.26 enabling us to classify it as a candidate planet orbiting the star. Although the significance of the signal corresponding to this candidate was only barely above the threshold of $\alpha = 20.51$, it was detected according to all our criteria and we did not identify any counterparts in the activity indicators suggestive of stellar rather than planetary origin. HD 150706 thus serves as an example of a previously undetected albeit significant signal that is barely detected in the data but can still be classified as a candidate planet according to our criteria. We have plotted the phase-folded radial velocities of HD 150706 in Fig. \ref{fig:HD150706_signal} for visual inspection. We note that although the signal appears slightly eccentric in Fig. \ref{fig:HD150706_signal}, the eccentricity is not statistically significantly different from zero. No additional signals were detected in the HD 150706 Keck radial velocities.

\begin{table*}
[b!]
\caption{Signals in the Keck data sets with likelihood ratio $\Delta \ln L_{j} = \ln L(j=k) - \ln L(j=k-1)$ between models with $k$ and $k-1$ signals for which the ratio exceeds a value corresponding to a 0.1\% FAP threshold of a three-parameter model (amplitude $K$, period $P$, and phase $\phi$).
(\textit{Sample: full table in electronic version})}
\vskip 0.1in
\label{tab:signals}
\begin{minipage}{\textwidth}
\begin{center}
\begin{tabular}{lccccll}
\hline \hline
Target & $n_{signals} $ & $\Delta \ln L_{k}$ & $P$ & $K$ & Interpretation & Notes\\
 &  &  & (days) & (ms$^{-1}$) \\ 
\hline
BD-103166 &	1 	&	73.71	&	3.4879$\pm$0.0001		&	58.56$\pm$1.46	&	Published	&		\\
GL83.1	&	1	&	17.74	&	771.9989$\pm$12.9944	&	16.38$\pm$2.87	&	SRC		&		\\
GL317	&	1	&	95.44	&	696.4070$\pm$0.9607	&	72.95$\pm$1.01	&	Published	&		\\
GL317	&	2	&	50.26	&	4725.5918$\pm$120.4710	 &	16.57$\pm$1.28	&	Candidate	&		\\
GL388	&	1	&	20.68	&	1.8080$\pm$0.0001		&	19.92$\pm$2.83	&	SRC		&		\\
GL686	&	1	&	20.45	&	15.5303$\pm$0.0030	&	3.46	$\pm$0.56	&	SRC		&		\\
GL687	&	1	&	47.05	&	38.1367$\pm$0.0130	&	5.66	$\pm$0.52	&	Published	&		\\
GL803	&	1	&	25.57	&	3.1052$\pm$0.0005		&	264.10$\pm$31.64	&	Activity	&	1	\\
GL876	&	1	&	312.79	&	61.0307$\pm$0.0010	&	211.03$\pm$0.63	&	Published	&		\\
GL876	&	2	&	407.69	&	30.2290$\pm$0.0006	&	88.60$\pm$0.60	&	Published	&		\\
GL876	&	3	&	185.22	&	15.0425$\pm$0.0007	&	19.09$\pm$0.55	&	Published	&	2	\\
GL876	&	4	&	104.94	&	1.9379$\pm$0.0000		&	5.93	$\pm$0.34	&	Published	&		\\
HD166	&	1	&	23.48	&	264.6946	$\pm$1.2329	&	35.29$\pm$3.14	&	Published	&		\\
HD1326	&	1	&	38.18	&	11.4424$\pm$0.0018	&	1.93	$\pm$0.21	&	Candidate	&		\\
HD1326	&	2	&	18.74	&	45.1422$\pm$0.0522	&	1.35	$\pm$0.23	&	Activity	&	3	\\
HD1388	&	1	&	31.52	&	5081.7874$\pm$99.1337	&	12.52$\pm$1.68	&	Candidate	&	4	\\
HD 38529  &     1      &      147.21      &      2135.95073$\pm$2.6032  &     162.61$\pm$1.18       &      Published  &    5       \\
\hline \hline
\end{tabular}
\vskip 0.1 in
\end{center}
\footnotesize{
Note 1: Signal coincides with a photometric period based on ASAS V-band photometry. \\ 
Note 2: Spurious signal that emerges from the use of Keplerian fitting functions for a system the requires an N-body model. \\ 
Note 3: A corresponding signal in the S-index data at a very nearby period, is likely indicative of the star's P$_{rot}$ \\
Note 4: Signal is not constrained in the parameter space (1 day \textless\ P \textless\ T$_{obs}$}) of the analyses. \\
Note 5: Signal coincides with the activity cycle in the S-index but is strong enough to be interpreted as a candidate planet. \\
Note 6: Likely artifact signal caused by orbital evolution of the system. \\
Note 7: Daily alias of the known transiting planet orbiting the star with an orbital period of 0.73 days \citep{winn2011}. \\
Note 8: The strongest S-index counterpart in our sample; signal is likely caused by activity. \\
\end{minipage}
\end{table*}

\begin{table*}
[b!]
\caption{Maximum likelihood parameter estimates and standard errors for the ``nuisance'' parameters.
(\textit{Sample: full table in electronic version})}\label{tab:noise_parameters.}
\vskip 0.1 in
\begin{minipage}{\textwidth}
\begin{center}
\begin{tabular}{lcccccc}
\hline \hline
Target & $\dot{\gamma}$ & $\sigma_{\rm J}$ & $c_{\rm S}$ & $\Delta \ln L_{\rm S}$ & $P_{\rm S}$ \\
 & (ms$^{-1}$year$^{-1}$) & (ms$^{-1}$) & (ms$^{-1}$) & & (days) \\ 
\hline
0748-01711-1 & 76.498$\pm$16.029 & 40.28$\pm$8.63 & 1904.802$\pm$1352.44 &  &  \\
BD-103166 & 0.069$\pm$0.499 & 4.67$\pm$0.94 & 152.058$\pm$48.888 & 7.909 & 430.097 \\
G60-06 & 0.256$\pm$0.627 & 4.9$\pm$1.54 & 32.809$\pm$28.534 &  &  \\
G097-054 & -0.116$\pm$0.572 & 6.14$\pm$2.15 & 11.982$\pm$14.361 &  &  \\
G161-29 & -0.447$\pm$1.281 & 13.39$\pm$3.85 & 21.394$\pm$118.851 &  &  \\
G192-13 & -0.705$\pm$0.789 & 11.66$\pm$2.53 & 20.646$\pm$11.467 &  &  \\
G195-59 & -0.298$\pm$0.781 & 5.96$\pm$2.64 & -0.686$\pm$3.093 &  &  \\
G205-028 & -0.183$\pm$1.352 & 5.61$\pm$2.82 & 8.763$\pm$7.76 &  &  \\
G207-019 & 1.75$\pm$1.86 & 5.97$\pm$2.75 & 6.111$\pm$17.322 &  &  \\
G244-047 & -0.386$\pm$0.647 & 6.8$\pm$2.49 & -8.785$\pm$13.874 &  &  \\
GL26 & 0.908$\pm$0.573 & 7.71$\pm$1.16 & -1.015$\pm$0.669 &  &  \\
GL47 & 0.373$\pm$0.945 & 8.1$\pm$2.82 & -24.605$\pm$21.704 &  &  \\
GL48 & 0.402$\pm$0.091 & 2.3$\pm$0.35 & -5.21$\pm$4.011 &  &  \\
GL49 & 0.25$\pm$0.17 & 2.47$\pm$0.99 & 1.303$\pm$3.248 &  &  \\
GL83.1 & 0.468$\pm$0.494 & 10.92$\pm$1.3 & -0.768$\pm$0.594 &  &  \\
GL87 & 0.409$\pm$0.109 & 2.75$\pm$0.38 & 8.732$\pm$7.979 &  &  \\
GL105B & 2.039$\pm$0.943 & 6.5$\pm$2.68 & -21.413$\pm$18.861 &  &  \\
GL107B & 0.915$\pm$1.251 & 5.77$\pm$2.4 & 17.005$\pm$20.201 &  &  \\
GL109 & -0.141$\pm$0.108 & 2.03$\pm$0.53 & 6.039$\pm$4.385 &  &  \\
GL226 & 0.319$\pm$0.112 & 2.4$\pm$0.55 & 6.629$\pm$2.775 &  &  \\
GL239 & 0.39$\pm$0.124 & 3.71$\pm$0.35 & 2.305$\pm$4.121 & 26.308 & 5835.988 \\
GL250B & 0.584$\pm$0.401 & 3.55$\pm$0.65 & -4.465$\pm$4.676 &  &  \\
GL273 & 0.098$\pm$0.101 & 2.63$\pm$0.36 & 0.433$\pm$3.582 & 9.598 & 5868.975 \\
GL317 & 3.449$\pm$0.439 & 3.47$\pm$0.71 & -0.871$\pm$1.325 &  &  \\
GL357 & -0.066$\pm$0.173 & 3.38$\pm$0.69 & 11.814$\pm$11.279 &  &  \\
GL382 & 0.348$\pm$0.14 & 4.45$\pm$0.37 & 5.286$\pm$3.208 & 9.738 & 1.033 \\
GL388 & -1.407$\pm$1.418 & 11.71$\pm$1.58 & 1.196$\pm$0.725 & 8.801 & 148.538 \\
GL393 & 0.342$\pm$0.13 & 3.31$\pm$0.39 & -0.283$\pm$3.791 &  &  \\
GL397 & 2.625$\pm$0.356 & 7.55$\pm$1.15 & 20.352$\pm$7.881 &  &  \\
GL406 & -0.212$\pm$0.625 & 12.95$\pm$1.98 & -0.287$\pm$0.231 &  &  \\
GL408 & 0.515$\pm$0.139 & 3.84$\pm$0.44 & -0.72$\pm$5.413 & 9.764 & 217.618 \\
GL412A & 0.089$\pm$0.077 & 2.03$\pm$0.19 & -0.981$\pm$5.15 & 18.209 & 28.486 \\
\hline \hline
\end{tabular}
\end{center}
\end{minipage}
\end{table*}
\vskip 0.1in

\section{Discussion of selected planet candidates cases}

\subsection{HD 68017}

HD 68017 is a known double star with an M5V companion at a projected separation of 13 AU \citep{crepp2012}. Given that we limited our signal searches for periods up to the data baseline, the 5844-day signal that we observe in the Keck radial velocities is likely just a lower bound estimate for the orbital period of this companion, although we cannot rule out a massive long-period planet orbiting the star either. Moreover, some of the apparent long-period radial velocity variability could be connected to a magnetic activity cycle of the star that we observe in the S-indices at a period of 5554 days. This period is also likely to be an estimate for a lower bound due to our choice to limit signal searches by the data baseline.

\subsection{HD 75732}

Although HD 75732 (55 Cancri) is reportedly a 5-planet system \citep{nelson2014}, we could only obtain evidence for 4 signals with the Keck data alone. However, due to the fact that we limited the period space of our signal searches to the interval [$P_{\rm min}, T_{\rm obs}$], where $P_{\rm min} = 1$ days, the fourth signal that we detected in the Keck data at a period of 2.80 days corresponds to the daily alias of the known transiting super-Earth with an orbital period of 0.74 days \citep{winn2011}.

\begin{figure}
\center
\includegraphics[angle=270, width=0.80\textwidth,clip]{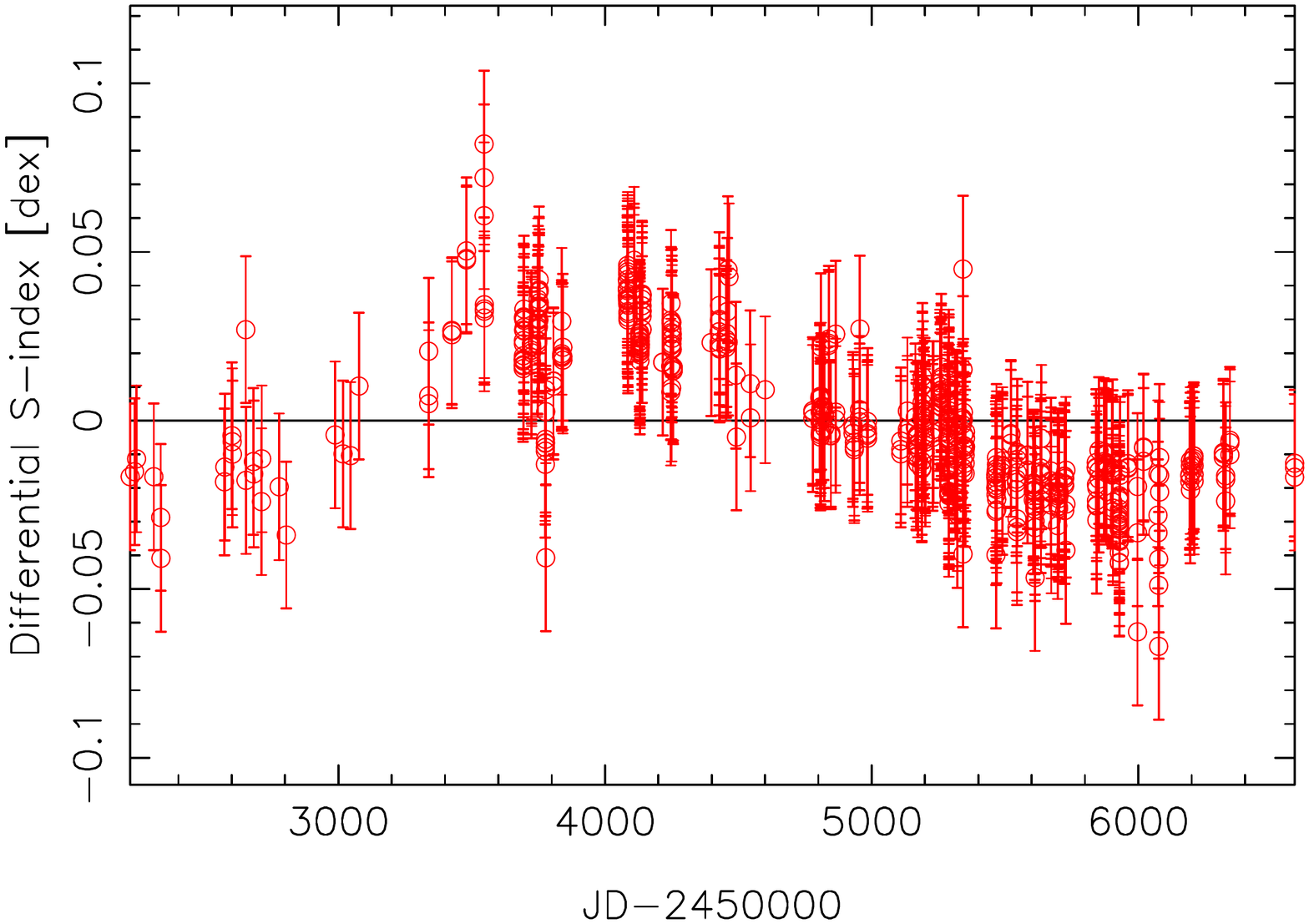}
\includegraphics[angle=270, width=0.80\textwidth,clip]{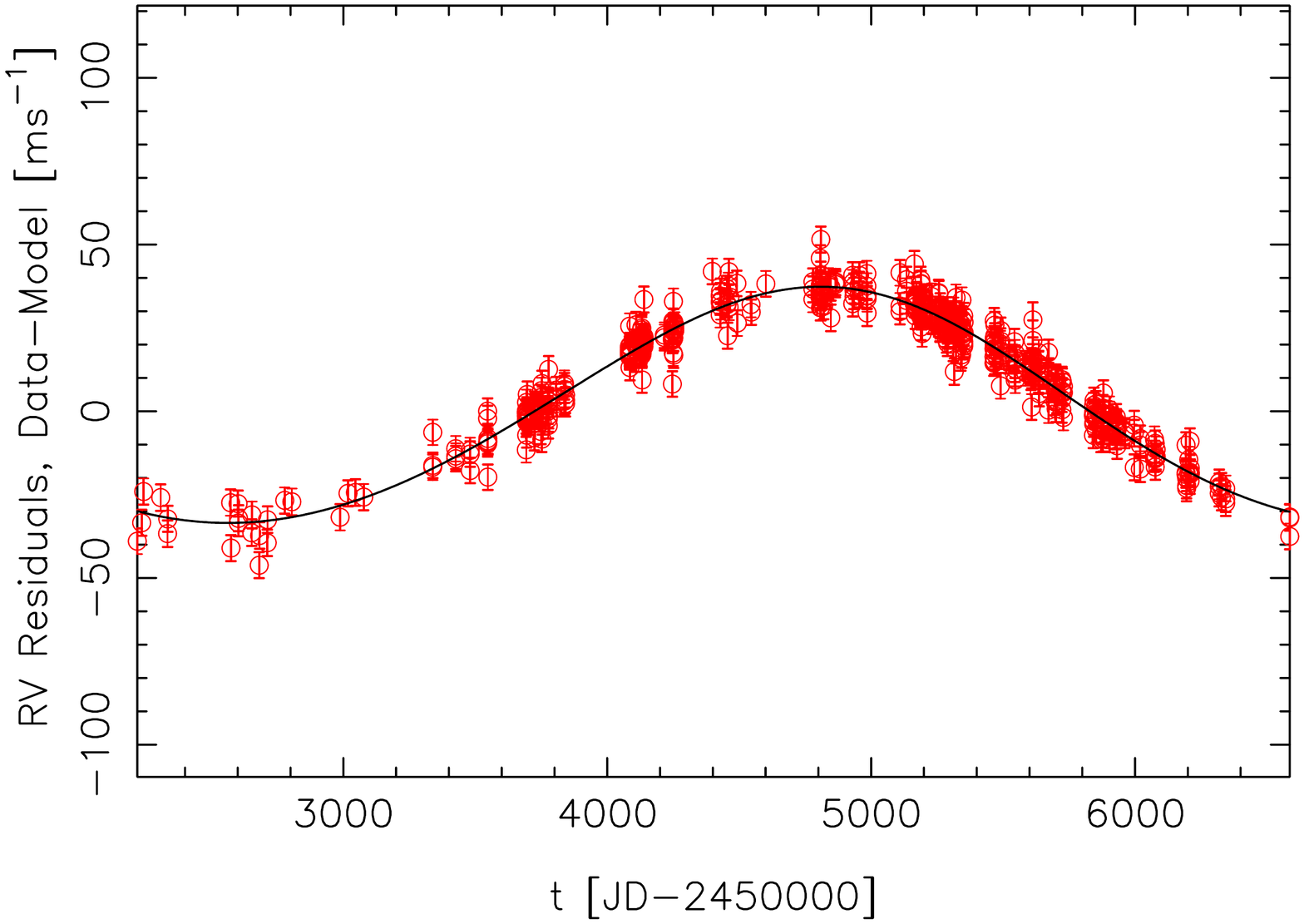}
\caption{Differential S-index variability for HD 75732 with respect to the data mean (top panel) and the radial velocities with the three shortest periodicities subtracted (bottom). In the bottom panel, the black curve denotes the modeled Keplerian signal.}\label{fig:HD75732_signals}
\end{figure}

 \citet{wright2015} point out that the longest periodicity in the data is similar to the period of the stellar activity variation, but notes that the two time series are out of phase, implying a coincidence between the durations of the orbital period and the stellar activity cycle rather than a connection. Moreover, though the variations in the S-indices appear to trace the radial velocity variability, this is not verified statistically. When accounting for the four Keplerian signals in the data, the radial velocities are independent of the S-indices with parameter $c_{\rm S}$ estimated to have a value of -27.3$\pm$19.2, which is not significantly different from zero with a 2-$\sigma$ level. If the longest periodicity was connected to the variations in the S-index and thus stellar activity, we would not expect the signal to be modeled well with a Keplerian periodicity. Moreover, the phases of the two signals do not coincide (Fig. \ref{fig:HD75732_signals}) leaving us without any evidence supporting the activity-induced origin for the longest radial velocity periodicity apart from a coincidence in the period that mostly arises from the fact that the periods of the two signals are constrained from above by the baseline of the data.

\subsection{HD 95735}

\begin{figure}
\center
\includegraphics[angle=270, width=0.80\textwidth,clip]{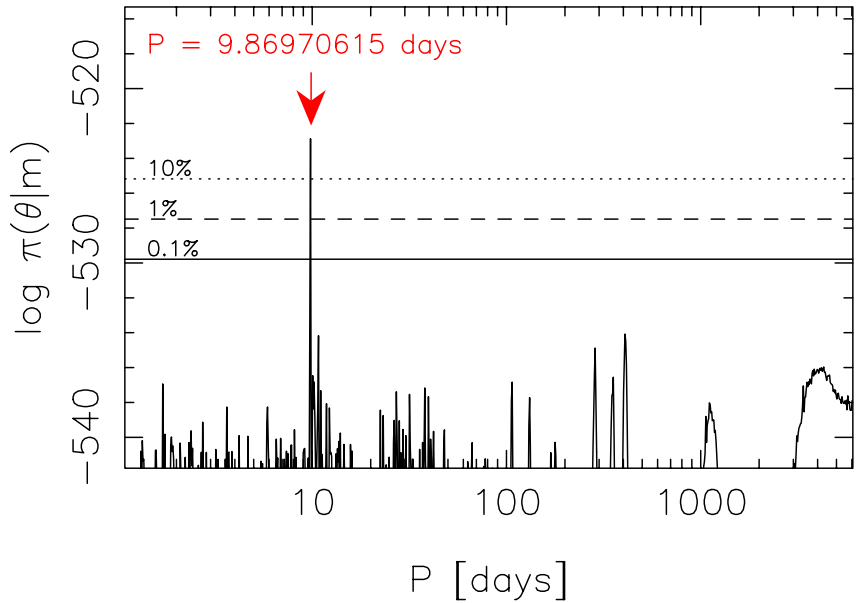}
\caption{Estimated posterior probability density as a function of the period parameter of the signal in a one-Keplerian model given radial velocities of HD 95735. Red arrow denotes the global probability maximum and the horizontal lines correspond to equiprobability contours at 10\% (dotted), 1\% (dashed), and 0.1\% (solid) of the maximum.}\label{fig:HD95735_posterior}
\end{figure}

The radial velocity measurements for HD 95735 (GJ 411, Lalande 21185) support the existence of a planet candidate orbiting the star with an orbital period of 9.8693$\pm$0.0016 days. The corresponding Keplerian signal has an amplitude, $K=1.90\pm$0.31 ms$^{-1}$ and it occupies a unique maximum in the estimated posterior probability density of the one-Keplerian model (Fig. \ref{fig:HD95735_posterior}). With a Hipparcos parallax of 393.42$\pm$0.70 mas (8.3 light years)  \citep{vanleeuwen2007} the star is the 4th closest main sequence stellar system to the Sun after the $\alpha$ Centauri triple system, Barnard's star, and Wolf 359. The planetary candidate has a minimum mass, $M\sin(i)=3.8 M_{\oplus}$, and receives an energy flux from the star roughly 5.3$\times$ larger than the flux received by Earth from the Sun (based on $R_{\star}=0.393 R_{\odot}$, $T_{\rm eff}=3828$, and $M_{\star}=0.46\,M_{\odot}$). The {\it a-priori} geometric probability that the planet can be observed in transit is $P=2.6\%$. The bright (V=7.5) apparent magnitude of HD 95735 coupled with the modest stellar radius suggest that a ground-based photometric monitoring campaign for this system would be worthwhile.

\begin{figure}
\center
\includegraphics[angle=270, width=0.80\textwidth,clip]{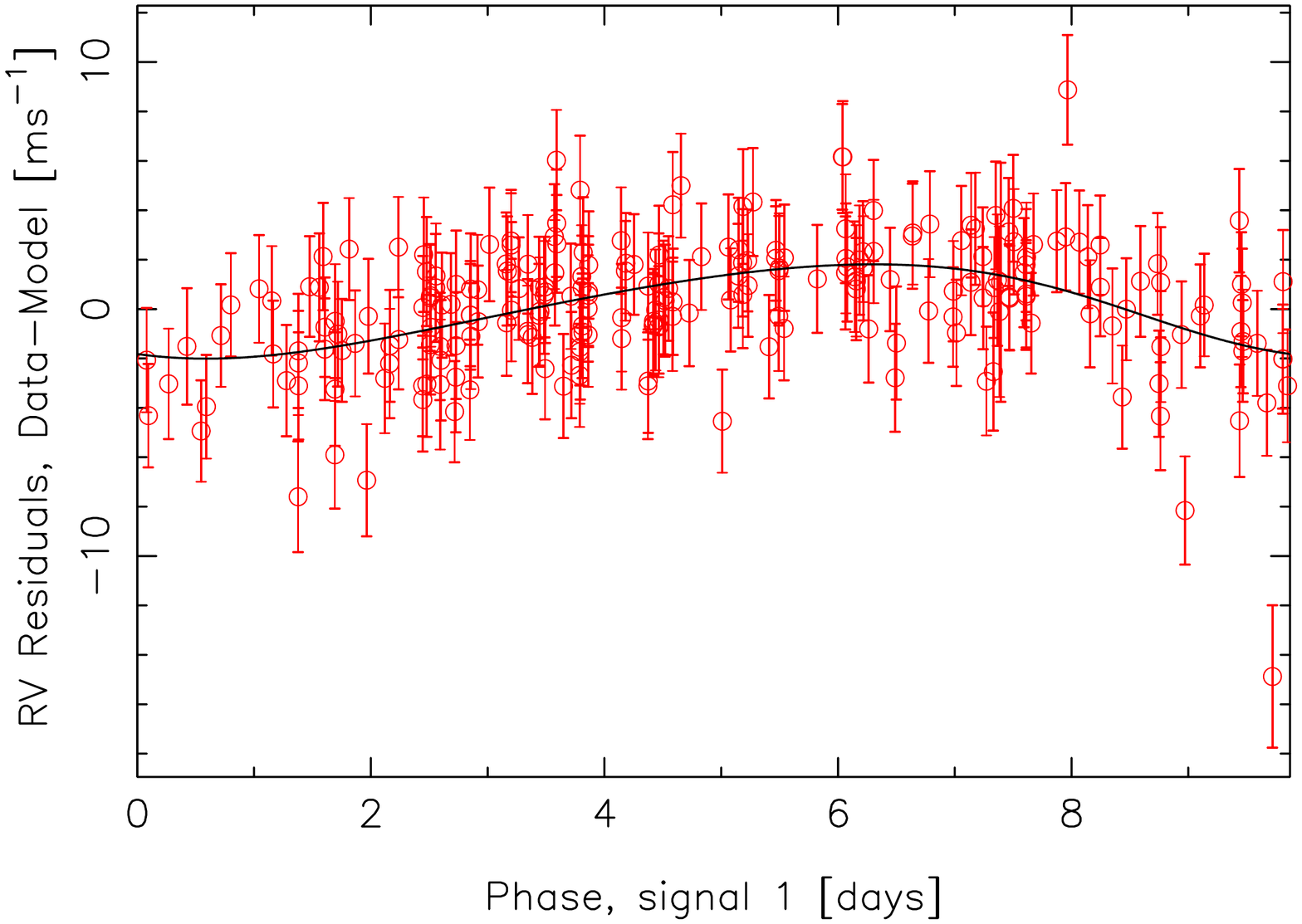}

\includegraphics[angle=270, width=0.80\textwidth,clip]{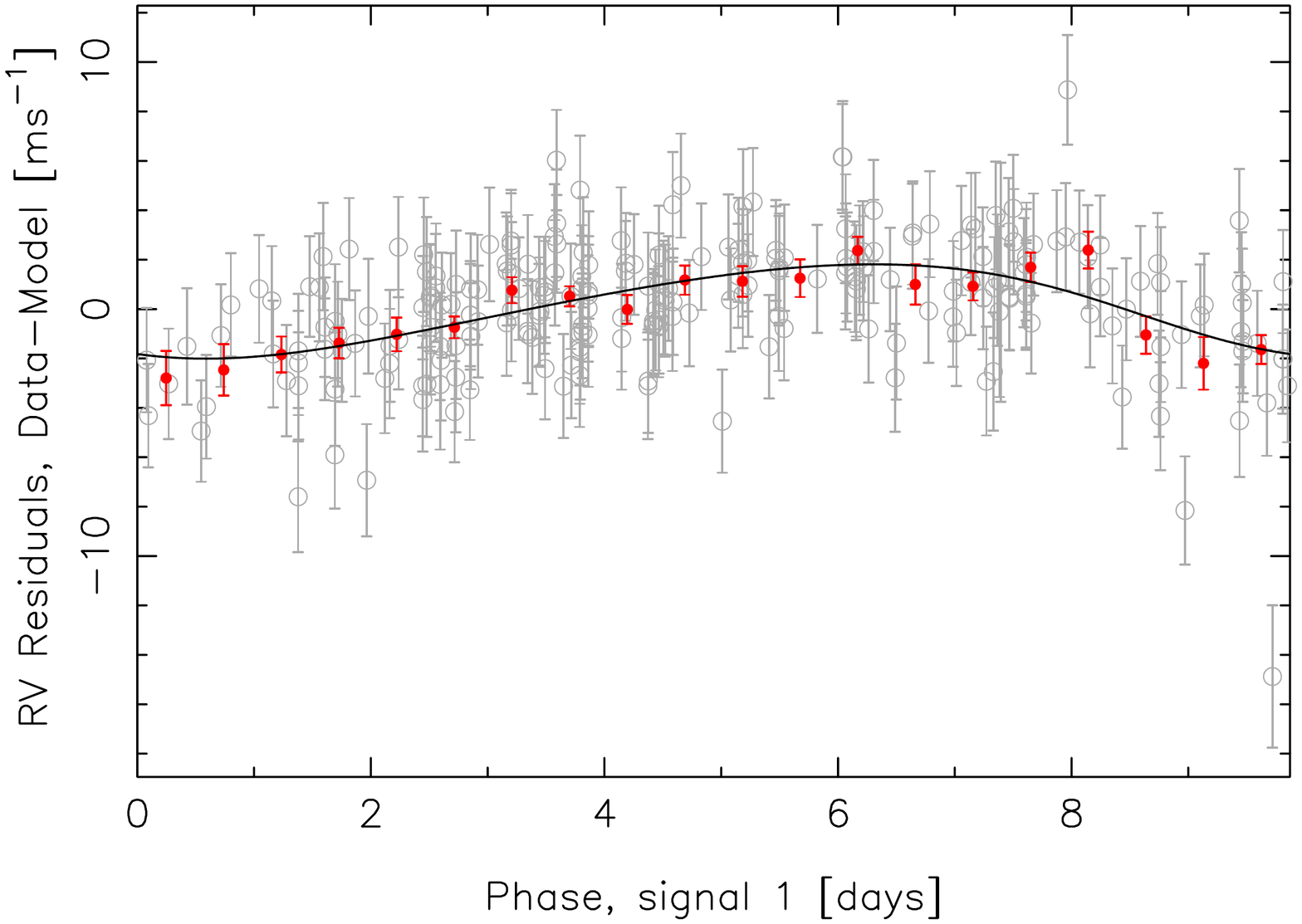}
\caption{Keck radial velocities for HD~95735 folded on the phase of the detected signal at a period of 9.87 days. In the bottom panel, weighted averages (red filled circles) are shown on top of the velocities (grey circles) when dividing the orbital phase into 30 bins. The solid curve denotes the modeled Keplerian radial velocity variability.}\label{fig:HD95735_phased}
\end{figure}

The radial velocity signal in the HD 95735 data is well-constrained and satisfies all our signal detection criteria with $\Delta \ln L_{1} =$ 22.24, which is in excess of the detection threshold we required to interpret a signal as a candidate planet ($\alpha = 20.52$). We have plotted the phase-folded radial velocities of HD 95735 in Fig. \ref{fig:HD95735_phased} for visual inspection.

\subsection{HD 154345}

\citet{wright2015} suggests that the radial velocity signal from this star, which was initially reported as a planetary detection by \citep{wright2008}, is better interpreted as a stellar activity cycle as a consequence of the apparently high correlation (that has persisted over a much longer time base since the initial discovery) between the RV and S-index variations. We find, however, no significant linear correlation between RVs and the S-index. In Table 3, the parameter, $c_{\rm S}$,  quantifying this dependence of RV on S-index has a value of 6.9$\pm$21.4, implying that it is not significantly different from zero and, by extension, that the radial velocity variations are thus not connected to the S-indices. Therefore, despite the visual similarity of the RV and S-index curves, we find no evidence that the RV variations are linearly connected to the S-index. The only connection between them is an apparent period coincidence and phase coincidence observed by \citet{wright2015}. Moreover, the signal in the S-indices appears to be quasiperiodic as would be expected from a signal connected to a stellar magnetic cycle whereas the radial velocity signal with a period of 3296$\pm$0.44 days does not seem to change as a function of time.

\begin{figure}
\center
\includegraphics[angle=270, width=0.80\textwidth,clip]{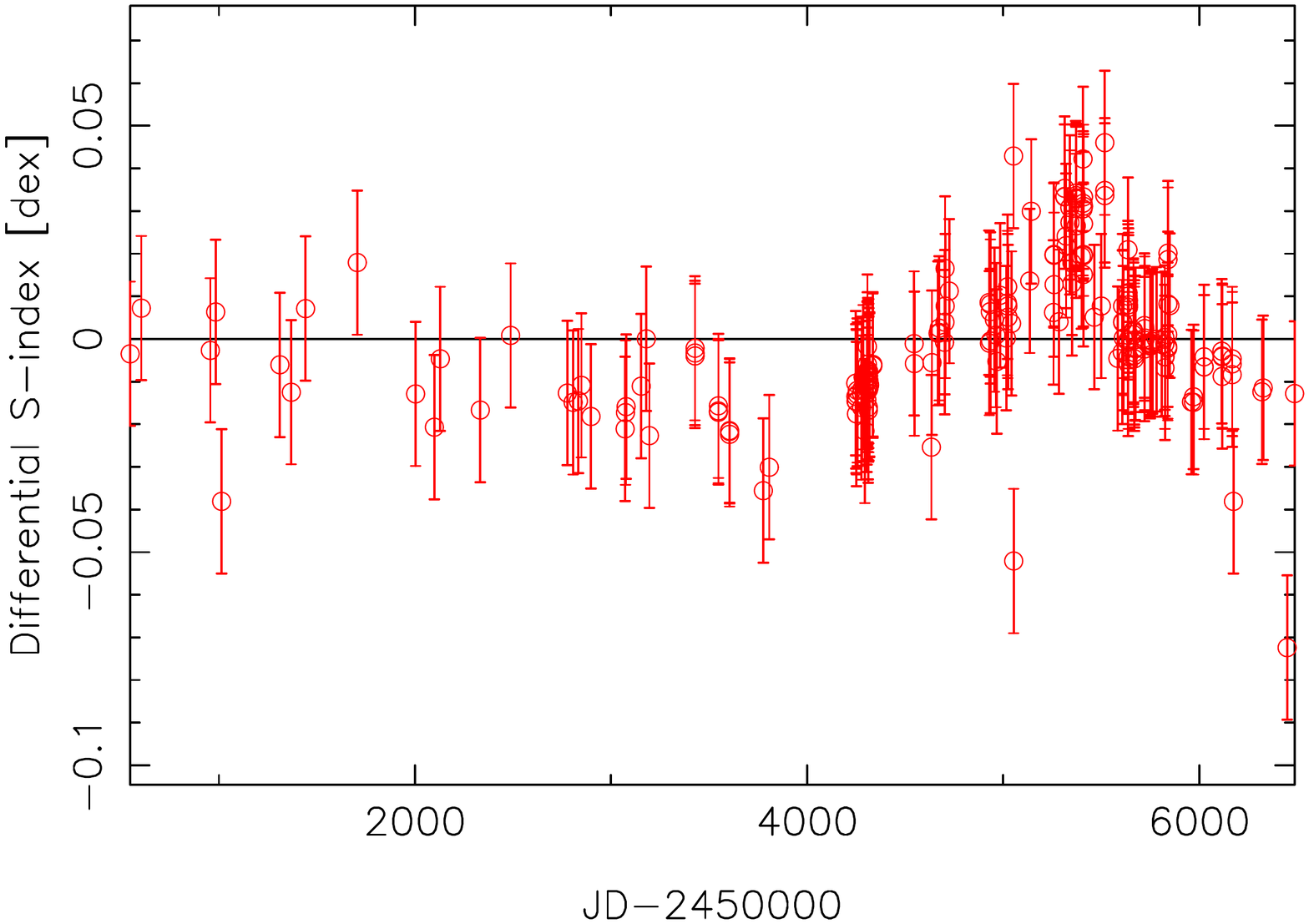}

\includegraphics[angle=270, width=0.80\textwidth,clip]{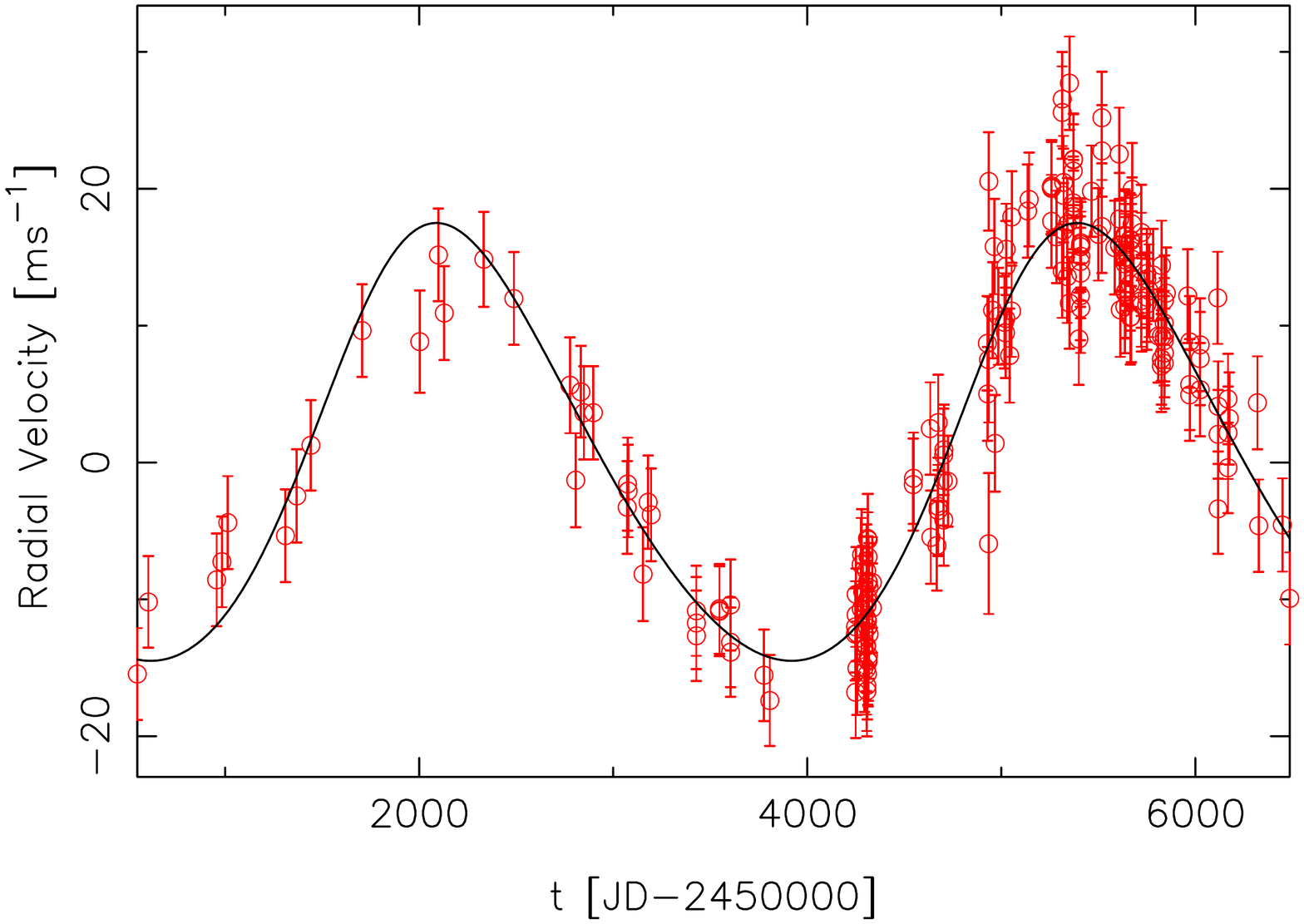}
\caption{S-index time series of HD 154345 with respect to the mean (top panel) and the radial velocity time-series with the Keplerian curve overplotted (bottom panel).}\label{fig:HD154345_plots}
\end{figure}

We illustrate the qualitative difference between the radial velocity and S-index variability in Fig. \ref{fig:HD154345_plots}, where we have plotted the radial velocity and S-index time-series. Although there are coincidental variations\footnote{This coincidence can be contrasted with the example in our own system that the orbital period of Jupiter is approximately equal to the Solar magnetic cycle.}, in particular between JDs 2455000-2456000, the overall shape of the variability is different and the two time-series are therefore not correlated when accounting for the signal of the planetary companion. It is thus our interpretation that the signal in the HD 154345 radial velocities is likely to be caused by a planet candidate analogous to our own Jupiter.

\subsection{HD 156668}

The radial velocities of HD 156668 point towards a system with at least two planet candidates. With more data, we can confirm the published 4.64-day periodicity \citep{howard2011} and likewise interpret it as a signal of a planet candidate. The Keck data now show no evidence for the hypothesis that the  4.64-day periodicity is an alias of the 1.27-day true period \citet{DF10}. Moreover, according to our analyses, there is another planet candidate orbiting the star with an orbital period of 855$\pm$23 days. We also find evidence in the S-indices for the stellar magnetic cycle with a period of 3753 days but this cycle has no counterpart in the radial velocity data, and is unlikely to weaken an interpretation that the 855-day periodicity arises from the Keplerian motion of a planet.
\subsection{HD 185144}

The HD 185144 radial velocities contain a low amplitude ($K = $1.39$\pm$0.23 ms$^{-1}$) signal with a period of 2644$\pm$197 days. This signal, however, appears to coincide with a magnetic cycle of the star with a period of 2184 days that we observe in the S-indices. Due to the low amplitude of the signal in the radial velocities and its proximity to the magnetic cycle, we cannot conclude that the signal is caused by a planet orbiting the star rather than activity. We thus interpret the periodicity as a radial velocity counterpart of the stellar magnetic cycle. In the future, frequent monitoring of the radial velocities and CaII H\&K line emissions of the star should permit assessment of whether the similarity in periods is a coincidence rather than a causal relationship.

\subsection{HD 207832}

\citet{hagh2012} reported two Jovian-type planets orbiting HD 208732. We had no difficulties in detecting the signals corresponding to these planet candidates in the Keck data. We find evidence, however, for a stellar magnetic cycle in the S-indices at a period of 1369 days. This is reasonably close to the orbital period of the outer candidate of 1252$\pm$65 days. However, the radial velocity signal is explained much better by a Keplerian model than simple linear relationship between the radial velocities and S-indices -- this is apparent because when modeling the candidates reported by \citet{hagh2012} there is no evidence for a dependence of the radial velocities on the S-indices in the sense that parameter $c_{\rm S}$ in Eq. (\ref{eq:model}) would be statistically significantly different from zero.

\subsection{HD 265866}

\begin{figure}
\center
\includegraphics[angle=270, width=0.80\textwidth,clip]{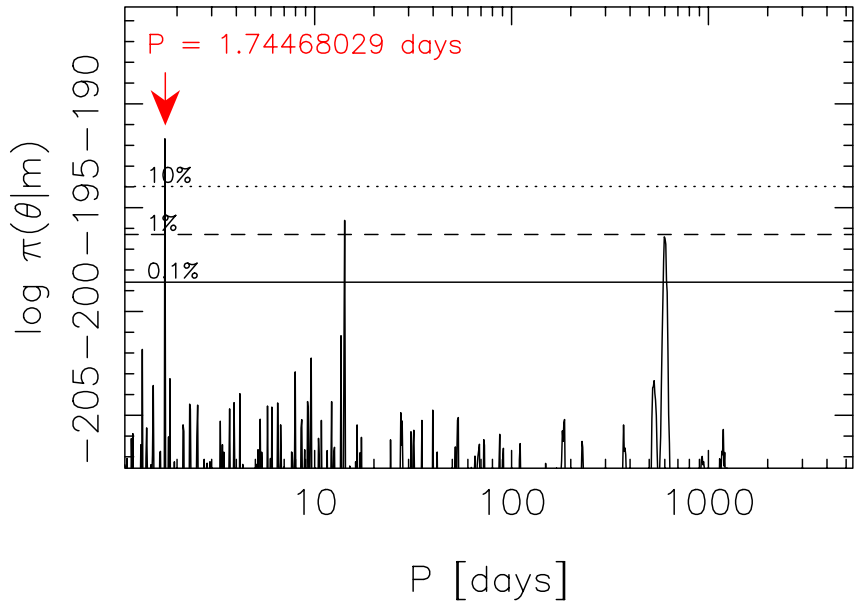}
\caption{Estimated posterior probability density as a function of the period parameter of the signal in a one-Keplerian model given radial velocities of HD 265866. Red arrow denotes the global probability maximum and the horizontal lines correspond to equiprobability contours at 10\% (dotted), 1\% (dashed), and 0.1\% (solid) of the maximum.}\label{fig:HD265866_posterior}
\end{figure}

\begin{figure}
\center
\includegraphics[angle=270, width=0.80\textwidth,clip]{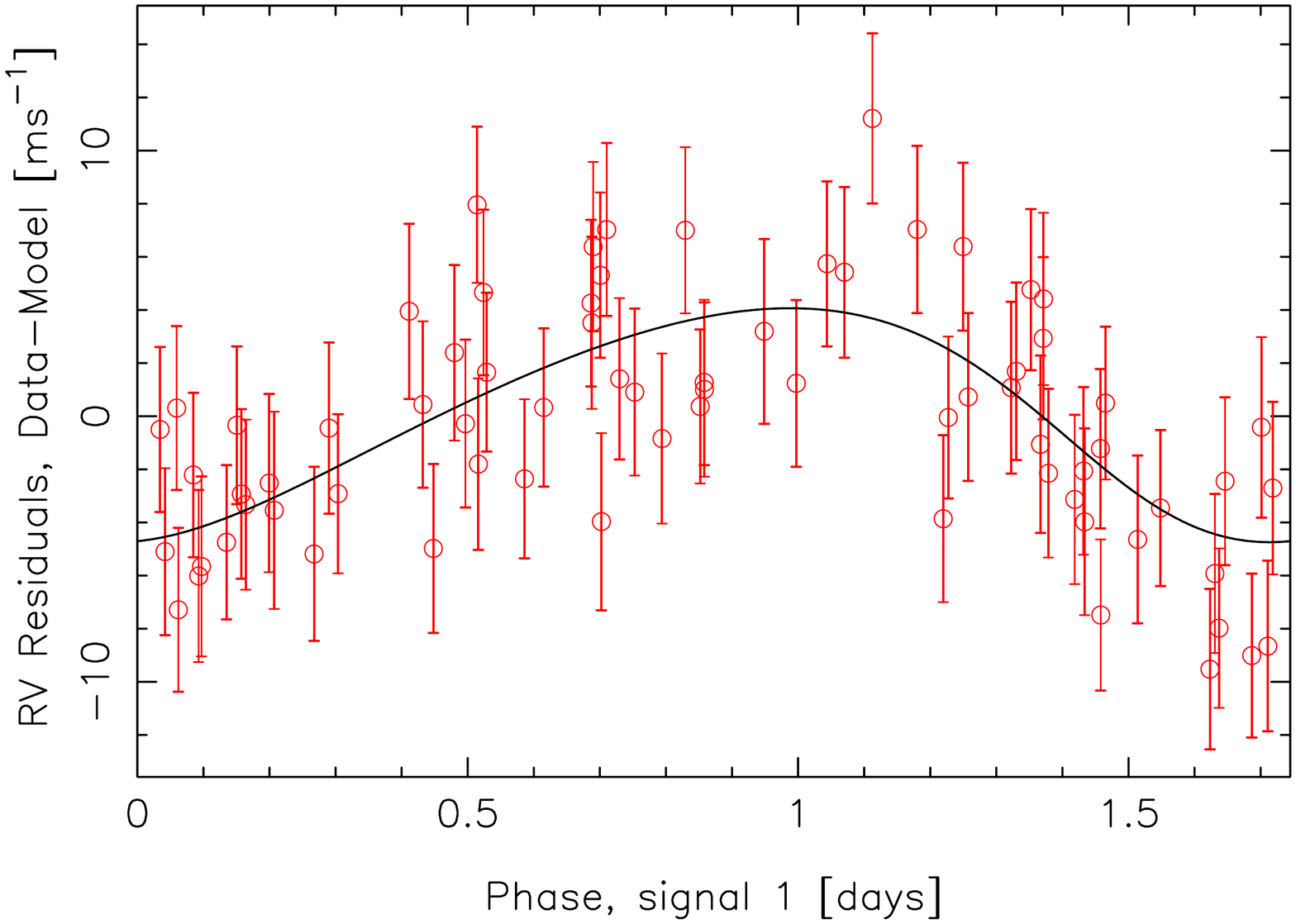}
\caption{Radial velocities of HD 265866 folded of the phase of the detected signal.}\label{fig:HD265866_signal}
\end{figure}

The radial velocities of the nearby M dwarf HD 265866 (GJ 251, HIP 33226) showed evidence in favor of two signals that we interpret are being caused by candidate planets orbiting the star. Our searches for a signal in the data with a one-Keplerian model identified a strong signal at a period of 1.74471$\pm$0.00005 days with an amplitude of 3.97$\pm$0.64 m$^{-1}$ (Fig. \ref{fig:HD265866_posterior}). We note that in Fig. \ref{fig:HD265866_posterior}, there are two local maxima in the period space at periods of 14 and 600 days, respectively, suggestive of alternative solutions or additional signals in the data. However, the 1.74-day signal appears to explain the variations in the data the most convincingly and we thus adopt this periodicity as our preferred Keplerian periodicity. We have plotted the radial velocities folded on the phase of the signal in Fig. \ref{fig:HD265866_signal}. We strongly encourage future monitoring of this star to assess the nature of the observed local probability maxima shown in Fig. \ref{fig:HD265866_posterior}.

\section{Discussion}

HIRES on Keck-I was a transformatively successful facility during the first decade of exoplanetary detection and characterization. In recent years, however, as thousands of additional exoplanets and planetary candidate detections have streamed in from other telescopes, both on the ground and in space, two factors have limited HIRES' effectiveness: 1) lack of adequate observing cadence, and 2) inadequate spectral resolution. The Keck telescopes are heavily-shared facilities that are also scheduled around the lunar calendar, and even the most successful planet detection consortia have rarely obtained more than a few dozen nights of time on HIRES per year. Furthermore, the lunation-driven scheduling makes it very hard for bright-time observers to obtain full phase coverage for planets that have orbital periods near integral multiples of the lunar month. When one considers the most interesting cases of very low mass, potentially rocky terrestrial planets and super-Earths, it is now known from {\it Kepler} that a large fraction of such exoplanet systems typically harbor multiple low-mass planets \citep{Batalha2013}. When confronted with a signal containing several low-amplitude periodicities, it is difficult to accurately measure orbital parameters without sufficient cadence. The problem is compounded by jitter stemming from stellar activity. As a consequence, the data presented here likely contain a plethora of bona-fide signals that lie beneath our noise thresholds. The velocity measurements presented in Table 1 will thus usefully support follow-on efforts that can access higher cadence and comparable (or improved) Doppler precision. An excellent recent example of this type of synergy was provided by the complex system of six planets around the star HD 219134 \citep{Motalebi15, Vogt15}. We amassed 331 precision velocities of this star with HIRES/Keck over seven years, and we were aware of at least 3 significant periodicities in this system as early as 2010. With the limited observing cadence achievable at Keck, however, we were unable to adequately parameterize this complex 6-planet system until 2015, after we had obtained 101 additional higher-cadence observations over two years with the Automated Planet Finder at Lick Observatory \citep{vogt2014a}.

Speed and precision comparisons are sometimes drawn between the ``iodine-based'' technique employed with HIRES, and the super-stabilized, fiber-scrambled technique of HARPS and its similar cousins. The HARPS and HARPS-variants approach utilizes a dual-white-pupil spectrometer, with stabilization of the instrument's line spread function provided by a fiber-feed image scrambler (or dual-scrambler), and spectrometer stabilization accomplished by enclosing the spectrometer in vacuum, and/or in a highly temperature-stabilized chamber. The HARPS has distinct advantages. With no need for iodine lines as a wavelength reference, HARPS does not have to acquire a separate high S/N template of each target star, and it circumvents the complex deconvolution issues that must be solved to adequately characterize the instrument line spread function when using the iodine technique. In addition, the HARPS approach is not restricted to the relatively narrow ($\Delta \lambda \sim 120$nm wavelength region of iodine, and can expand the wavelength range to typically 380-690 nm, generating 2.6 times more spectral line information. On the other hand, however, the HARPS approach is confronted by challenging wavelength calibration stability issues that are compounded over long time base lines. In summary, experience over the past 20 years has shown that \textit{both} techniques, when properly implemented, are capable of reaching sub \ms\ precisions, albeit each with its own attendant difficulties.

Direct comparisons of the ``iodine-technique" vs. the ``fiber-scrambled HARPS" approach, that cite HIRES as a reference example for the former, are somewhat misleading. HIRES was neither specifically designed for, nor optimized for, extreme precision RV work, and certainly, HIRES possesses limited effectiveness in connection with the highest precision radial velocity work primarily as a result of its inadequate spectral resolution. The highest resolution of HIRES (with scientifically effective {\it Throughput} at the slit) was set at $\sim$60,000 to allow general-purpose high resolution targeting of point sources as faint as V=22. Resolutions of typically greater than 110,000 are required, along with adequate pixel sampling, to reach sub \ms\ levels of precision. HIRES's resolution is almost a factor of two lower than the value $R\sim$115,000 of HARPS, a problem compounded by HIRES's lack of thermal stabilizion.

A substantially more useful comparison with HARPS can be made using Magellan's Planet Finder Spectrometer (PFS) \citep{Crane10}. PFS is a thermally stabilized slit-fed spectrometer that uses the classic iodine approach for wavelength reference and PSF determination. Like HIRES, PFS also is fed by a long slit and uses iodine rather than a fiber-scrambler for PSF stabilization. With a somewhat higher spectral resolution (80,000) than HIRES, PFS routinely achieves 1.1 \ms\ mean internal uncertainties per observation. Even $R\sim$80,000, however, is sub-optimal. PFS would benefit from an additional 45\% increase in its spectral resolution to bring it up to the 115,000 resolution of HARPS.  The new APF facility's Levy spectrometer \citep{vogt2014a} is very similar in optical design to PFS, but with the advantage of a somewhat higher $R\sim108,000$ resolution. The higher resolution enables the APF/Levy facility to achieve the same level of RV precision as HIRES, with about a factor of 6 fewer photons on M dwarfs. For M dwarf stars down to at least V=10, the APF facility achieves very comparable speed-on-sky as Keck, due largely to the higher spectral resolution of its spectrometer.

We are currently using APF to follow up on many promising targets from our 20-year HIRES exoplanet survey with much higher cadence \citep{burt2014, burt2015}. In particular, combining the higher cadence available from APF with the long time baseline of the HIRES program is proving to be very effective for characterizing complex multi-planet systems such as HD 141399 and HD 219134 \citep{vogt2014b, vogt2015}. We are also using PFS (in the south) and APF (in the north), and in combination over the declination overlap region, for K2 follow-up, and in the near future for TESS follow-up.

\section{Conclusion}

In this paper, we present 60,949 precision radial velocities of 1,624 stars obtained over the past 20 years from the LCES survey with the HIRES spectrometer on the Keck-I telescope. We also present a list of 117 unpublished likely planet-candidate signals gleaned from these data that appear to stem neither from stellar rotation nor from stellar activity. Given the automated and by extension, fully standardized signal searches performed in the current work, we can now study the global occurrence rates of planets orbiting the stars in the Keck sample. This requires the estimation of the detection probability function for the sample \citep{tuomi2014}. We are performing the required computations and expect to report them in a follow-up paper.

It is hoped that this unique corpus of radial velocities, by and of themselves, and especially when combined with additional RV and photometric data from other facilities, will be a valuable resource to the astronomical community for furthering the discovery and characterization of the galactic planetary census.

\acknowledgments

We gratefully acknowledge the long assistance of former California Planet Search colleagues Geoff Marcy, Debra Fischer, Jason Wright and Katie Peek for their
help in tending many nights at the Keck I telescope. SSV gratefully acknowledges support from NSF grants AST-0307493 and AST-0908870. SSV also gratefully acknowledges support from NASA grant NNG04GE18G that partially funded the HIRES focal plane upgrade, and NASA grant NAG-4445 that funded the commissioning of the HIRES exposure meter. RPB gratefully acknowledges support from NASA OSS Grant NNX07AR40G, the NASA Keck PI program, and from the Carnegie Institution of Washington. The work herein is based on observations obtained at the W. M. Keck Observatory, which is operated jointly by the University of California and the California Institute of Technology, and we thank the UC-Keck and NASA-Keck Time Assignment Committees for their support. This research has made use of the Keck Observatory Archive (KOA), which is operated by the W. M. Keck Observatory and the NASA Exoplanet Science Institute (NExScI), under contract with the National Aeronautics and Space Administration.
We also wish to extend our special thanks
to those of Hawaiian ancestry on whose sacred mountain of Mauna Kea we are privileged to be guests. Without their generous hospitality, the Keck observations presented herein would not have been possible.


\clearpage
\appendix

\section{Statistical model}\label{sec:statistical_model}

The statistical model for radial velocities consists of five additive parts: superposition of $k$ Keplerian signals $f_{k}$, linear acceleration $\dot{\gamma}$, reference velocity $\gamma$, correlated noise component, activity component, and white noise. These can be modeled by a forward model of a radial velocity measurement $m_{i}$ obtained at time $t_{i}$ written as
\begin{equation}\label{eq:model}
  m_{i} = f_{k}(t_{i}) + \dot{\gamma} + \gamma + \phi \exp \Bigg\{ \frac{t_{i-1}-t_{i}}{\tau} \Bigg\}\epsilon_{i-1} + c \xi_{i} + \epsilon_{i} ,
\end{equation}
where $\phi$ quantifies the magnitude of the moving average component used to model correlated noise \citep{tuomi2013,tuomi2014} with exponential smoothing in a time-scale of $\tau$, parameter $c$ quantifies the linear dependence of the velocity on the activity index $\xi_{i} = S_{i} - N^{-1}\sum_{i=1}^{N}S_{i}$, where $S_{i}$ is the measured S-index based on CaII H\&K emission at $t_{i}$, and $\epsilon_{i}$ is a Gaussian random variable with a zero mean and a variance of $\sigma_{i}^{2} + \sigma_{J}^{2}$ where $\sigma_{J}$ is also a free parameter of the model. $N$ denotes the number of measurements. We fixed the exponential smoothing time-scale to $\tau = 4$ days because this parameter is typically not well-constrained \citep{tuomi2014} as long as it accounts for correlations between nearby epochs arising from stellar activity and/or possible instrumental instability. This model has $5(k + 1)$ free parameters when $k$ Keplerian signals are included in the model.

This statistical model implies a likelihood function written as
\begin{equation}\label{eq:likelihood_model}
  L(m) = L(m_{1}) \cdot L(m_{2} | m_{1}) \cdots L(m_{N} | m_{N-1}). 
\end{equation}
For measurement $m_{i}$, the likelihood function is written as
\begin{equation}\label{eq:likelihood_model2}
  \log L(m_{i}) = -\frac{1}{2} \log \big( 2\pi \sigma_{i}^{2} \big) - \frac{R_{i}^{2}}{2 \sigma_{i}^{2}} ,
\end{equation}
where
\begin{equation}\label{eq:likelihood_model3}
  R_{i} = m_{i} - f_{k}(t_{i}) - \dot{\gamma} - \gamma - \phi \exp \Bigg\{ \frac{t_{i-1}-t_{i}}{\tau} \Bigg\}\epsilon_{i-1} - c \xi_{i}
\end{equation}

In addition to the likelihood function, we also used a prior probability density in our analyses. This prior was set uninformative such that $\pi(\theta_{i}) = 1$ for all free parameters $\theta_{i}$ except the orbital eccentricity. For eccentricity, we set $\pi(e) \propto \mathcal{N}(0, \sigma_{e}^{2})$ and $\sigma_{e} = 0.1$ \citep[see e.g.][]{tuomi2013c} to penalize high eccentricities in order to avoid situation where the Markov chains visit areas of high eccentricity in the parameter space disabling the reliable sampling of the period space in order to find signals. As indicated by the results in the current work, this choice did not prevent the detections of high-eccentricity solutions corresponding to eccentric giant planets. In the context of likelihood ratios that we used to assess the significances of the solutions, the eccentricity prior can be considered a regularization function, also known as Tikhonov-regularization \citep{tikhonov1977} and its application is thus justified even in the frequentist context.

When analyzing the time-series of the activity indices, we simplified the statistical model in Eq. \ref{eq:model} by setting the eccentricity of a signal equal to zero reducing the parameters of the signal to period, amplitude, and phase, and by setting $\phi = 0$ and $c = 0$ ms$^{-1}$.

\end{document}